\documentclass[11pt,a4paper]{article}
\usepackage[english]{babel}
\usepackage[utf8]{inputenc}
\usepackage{amsfonts}
\usepackage{amsmath}
\usepackage{amssymb}
\usepackage{mathtools}
\usepackage{graphicx}
\usepackage{bm}
\usepackage{xcolor}
\usepackage[font=small]{caption}
\usepackage{subcaption}
\usepackage{cite}
\def\linkcolor{cyan!70!black}
\usepackage[
colorlinks=true
,urlcolor=\linkcolor
,anchorcolor=\linkcolor
,citecolor=\linkcolor
,filecolor=\linkcolor
,linkcolor=\linkcolor
,menucolor=\linkcolor
,linktocpage=true
,pdfproducer=medialab
,pdfa=true
]{hyperref}
\usepackage{geometry}
\newcommand{\mathsym}[1]{{}}

\newcommand{\Sergey}[1]{{\color{green}#1}}

\DeclareUnicodeCharacter{2212}{-}
\pdfoutput=1
\input{tcilatex}

\begin{document}

\begin{titlepage}
 \begin{flushright}
     KANAZAWA-22-07
 \end{flushright}

\begin{center}
{\Large\bf Phenomenological and cosmological implications of\\a scotogenic three-loop neutrino mass model}

\vspace*{0.8cm}
Asmaa~Abada$^{a}$\footnote[1]{\href{mailto:asmaa.abada@ijclab.in2p3.fr}{asmaa.abada@ijclab.in2p3.fr}}, Nicol\'{a}s Bernal$^{b}$\footnote[2]{\href{mailto:nicolas.bernal@nyu.edu}{nicolas.bernal@nyu.edu}}, Antonio E. C\'{a}rcamo Hern\'{a}ndez$^{c,d,e}$\footnote[3]{\href{mailto:antonio.carcamo@usm.cl}{antonio.carcamo@usm.cl}},\\Sergey Kovalenko$^{d,e,f}$\footnote[4]{\href{mailto:sergey.kovalenko@unab.cl}{sergey.kovalenko@unab.cl}}, T\'{e}ssio B. de Melo$^{e,f,g}$\footnote[5]{\href{mailto:tessiomelo@gmail.com}{tessiomelo@gmail.com}}, and Takashi Toma$^{h,i}$\footnote[6]{\href{mailto:toma@staff.kanazawa-u.ac.jp}{toma@staff.kanazawa-u.ac.jp}}\\

\vspace*{.5cm}
$^{a}$P\^ole Th\'eorie, Laboratoire de Physique des 2 Infinis Ir\`ene Joliot Curie (UMR 9012)\\
CNRS/IN2P3, 15 Rue Georges Clemenceau, 91400 Orsay, France\\
$^{b}$New York University Abu Dhabi\\
PO Box 129188, Saadiyat Island, Abu Dhabi, United Arab Emirates\\
$^{{c}}$Universidad T\'{e}cnica Federico Santa Mar\'{\i}a, Casilla 110-V, Valpara\'{\i}so, Chile\\
$^{{d}}$Centro Cient\'{\i}fico-Tecnol\'{o}gico de Valpara\'{\i}so, Casilla 110-V, Valpara\'{\i}so, Chile\\
$^{{e}}$Millennium Institute for Subatomic Physics at the High-Energy Frontier, SAPHIR, Chile\\
$^{f}$ Departamento de Ciencias F\'isicas, Universidad Andr\'es Bello\\
Sazi\'e 2212, Piso 7, Santiago, Chile\\
$^{{g}}$Instituto de F\'{\i}sica e Matem\'{a}tica, Universidade Federal de Pelotas (UFPel)\\
Caixa Postal 354, CEP 96010-090, Pelotas, RS, Brazil\\
$^{{h}}$Institute of Liberal Arts and Science, Kanazawa University, Kanazawa 920-1192, Japan
$^{{i}}$Institute for Theoretical Physics, Kanazawa University, Kanazawa 920-1192, Japan

\vspace*{0.5cm}
\begin{abstract}
\noindent
We propose a scotogenic model for generating neutrino masses through a three-loop seesaw.
It is a minimally extended inert doublet model with a spontaneously broken global symmetry $U(1)'$ and a preserved $\mathbb{Z}_2$ symmetry.
The three-loop suppression allows the new particles to have masses at the TeV scale without fine-tuning the Yukawa couplings. The model leads to a rich phenomenology while satisfying all the current constraints imposed by neutrinoless double-beta decay, charged-lepton flavor violation, and electroweak precision observables. The relatively large Yukawa couplings lead to sizable rates for charged lepton flavor violation processes, well within future experimental reach.
The model could also successfully explain the $W$ mass anomaly and provides viable fermionic or scalar dark matter candidates.
\end{abstract}
\end{center}
\end{titlepage}

\section{Introduction}
Despite its remarkable consistency with the experimental data, the Standard Model (SM) fails to explain several issues, such as the observed SM fermion masses and mixing pattern, including the tiny light active neutrino masses, the current amount of dark matter (DM) relic density observed in the Universe, and the measured baryon asymmetry. Experiments with solar, atmospheric, and reactor neutrinos have provided evidence of neutrino oscillations caused by their non-vanishing masses. Several extensions of the SM have been constructed in order to explain the tiny masses of the light active neutrinos; in this work we consider neutrino masses generated radiatively; see \textit{e.g.} Ref.~\cite{Cai:2017jrq} for a review and Ref.~\cite{Arbelaez:2022ejo} for a comprehensive study of one-loop radiative neutrino mass models.

The simplest extension of the SM that provides masses for the light active neutrinos while keeping the SM gauge symmetry consists in adding right-handed (RH) Majorana neutrinos, singlets under the SM gauge symmetry, that mix with the active neutrinos, thus triggering the tree-level type-I seesaw mechanism at the origin of the tiny neutrino masses. This requires either extremely heavy RH Majorana neutrinos with masses close to the Grand Unification scale, or tiny Dirac neutrino Yukawa couplings if one considers TeV-scale RH Majorana neutrinos. In both scenarios, the mixing between active and heavy neutrinos is strongly suppressed, thus yielding very tiny rates for charged-lepton flavor-violating decays, several orders of magnitude below the current experimental sensitivity. This makes models with tree-level type-I seesaw realizations difficult to test via charged-lepton flavor-violating decays. In addition, those models are unable to successfully accommodate the current amount of DM relic density observed in the Universe. Alternatively, radiative seesaw models are examples of interesting and testable extensions of the SM explaining light neutrino masses and their mixing, where the seesaw mediators play an important role in successfully accommodating the current amount of DM relic density. In most radiative seesaw models, the light neutrino masses are generated at a one-loop level, thus implying that in order to successfully reproduce neutrino data as observed, one has to rely either on very small neutrino Yukawa couplings (of the order of the electron Yukawa coupling) or on an unnaturally small mass splitting between the CP-even and CP-odd components of the neutral scalar mediators.

Three-loop neutrino mass models have been proposed in the literature~\cite{Krauss:2002px, Aoki:2008av, Kajiyama:2013lja, Ahriche:2014cda, Ahriche:2014oda, Hatanaka:2014tba, Chen:2014ska, Jin:2014glp, Okada:2015hia, Nishiwaki:2015iqa, Ahriche:2015wha, CarcamoHernandez:2015hjp, Gu:2016xno, Cheung:2017efc, Dutta:2018qei, CarcamoHernandez:2019cbd, Cepedello:2020lul, Hernandez:2021iss} to provide a more natural explanation for the smallness of the light active neutrino masses than those relying on one- or two-loop seesaw realizations. However, most of these models have an important new particle content with, most of the time, one type of DM candidate, usually fermionic. In this work, we propose an extended Inert Doublet model (IDM) with moderate particle content, where light-active neutrino masses arise from a three-loop-level radiative seesaw mechanism. We extend the SM with a spontaneously broken global symmetry $U(1)'$ and a preserved $\mathbb{Z}_2$ symmetry that induces active neutrino masses at the three-loop level. Furthermore, the scalar sector is enlarged by the inclusion of four electrically neutral gauge singlet scalars, whereas the SM fermion content is augmented by the inclusion of two RH Majorana neutrinos. Due to the preserved $\mathbb{Z}_{2}$ symmetry, our model provides both fermionic and scalar DM candidates. The fact that in our model the active neutrino masses are generated at the three-loop level provides a more natural dynamical explanation for the tiny values of the light active neutrino masses compared to the usual one-loop-level radiative seesaw models. The model under consideration is consistent with the neutrino oscillation experimental data and allows us to successfully accommodate the measured DM relic abundance, as well as the constraints arising from charged lepton flavor violation (cLFV), oblique parameters $S$, $T$, and $U$, in addition to being consistent with the recently observed $W$-mass anomaly.
In this work, we address the phenomenological and cosmological implications of our model.  

The paper is organized as follows. In Section~\ref{Sec:model} we introduce the model, providing a detailed description of the scalar and fermionic sectors.
The generation of light neutrino Majorana masses is studied in Section~\ref{Sec:Neutrino}, with a discussion on the prediction of the effective neutrinoless double beta decay ($0\nu 2\beta$) effective mass. 
This model with its new field content and a spontaneously broken global symmetry $U(1)^{\prime}$ induces contributions to several observables; Section~\ref{sec:cLFVobs} is dedicated to electroweak precision observables and to charged lepton flavor violating processes (radiative $\ell\to\ell'\gamma$ and three-body decay $\ell\to \ell'\ell'\ell''$) and to the severe constraints the current bounds impose on the model. 
The domain wall problem is discussed in Section~\ref{Sec:cosmo}, as well as the possibility of successful scenarios for generating the baryon asymmetry of the Universe, and the viability of scalar and fermionic candidates for DM.  We summarize our findings in Section~\ref{Sec:conclusions}.
Expressions for the decay rates and other relevant amplitudes, as well as additional information on the analysis, are collected in the appendix.

\section{The model} \label{Sec:model}
We propose an extension of the inert doublet model (IDM)~\cite{Deshpande:1977rw, Ma:2006km, Barbieri:2006dq, LopezHonorez:2006gr, Belyaev:2016lok} where the SM gauge symmetry is extended by including a spontaneously broken global symmetry $U(1)'$ and a preserved discrete symmetry $\mathbb{Z}_2$.
The SM particle content is extended by four electrically neutral scalar singlets $\sigma$, $\rho$, $\varphi$, $\zeta$, and two RH neutrinos $N_{R_k}$ (with $k = 1$, 2). The preserved $\mathbb{Z}_2$ discrete symmetry allows for a stable scalar or fermionic DM candidate, corresponding to the lightest electrically neutral particle.
 The particle content of the model and the corresponding assignments under the $SU(3)_C \otimes SU(2)_L \otimes U(1)_Y \otimes U(1)' \otimes \mathbb{Z}_2$ symmetry are shown in Table~\ref{Themodel}. 
\begin{table}[hbt]
\renewcommand{\arraystretch}{1.3} \centering 
\begin{tabular}{|c||c|c|c|c|c|c|c|c|c|c|c|c|}
\hline
Field & $q_{iL}$ & $u_{iR}$ & $d_{iR}$ & $\ell_{iL}$ & $\ell_{iR}$ & $%
N_{R_{k}}$ & $\phi $ & $\eta $ & $\varphi $ & $\rho $ & $\zeta $ & $\sigma $
\\ \hline\hline
$SU(3)_{C}$ & $\mathbf{3}$ & $\mathbf{3}$ & $\mathbf{3}$ & $\mathbf{1}$ & $%
\mathbf{1}$ & $\mathbf{1}$ & $\mathbf{1}$ & $\mathbf{1}$ & $\mathbf{1}$ & $%
\mathbf{1}$ & $\mathbf{1}$ & $\mathbf{1}$ \\ \hline
$SU(2)_{L}$ & $\mathbf{2}$ & $\mathbf{1}$ & $\mathbf{1}$ & $\mathbf{2}$ & $%
\mathbf{1}$ & $\mathbf{1}$ & $\mathbf{2}$ & $\mathbf{2}$ & $\mathbf{1}$ & $%
\mathbf{1}$ & $\mathbf{1}$ & $\mathbf{1}$ \\ \hline
$U(1)_Y$ & $\frac{1}{6}$ & $\frac{2}{3}$ & $-\frac{1}{3}$ & $-\frac{1}{2}$
& $-1$ & $0$ & $\frac{1}{2}$ & $\frac{1}{2}$ & $0$ & $0$ & $0$ & $0$ \\ 
\hline
$U(1)'$ & $\frac{1}{3}$ & $\frac{1}{3}$ & $\frac{1}{3}$ & $-3$ & $-3$
& $0$ & $0$ & $3$ & $3$ & $-1$ & $0$ & $\frac{1}{2}$ \\ \hline
$\mathbb{Z}_2$ & $1$ & $1$ & $1$ & $1$ & $1$ & $-1$ & $1$ & $-1$ & $-1$ & $%
-1$ & $-1$ & $1$ \\ \hline
\end{tabular}
\caption{Particle charge assignments under the $SU(3)_C \otimes
SU(2)_L \otimes U(1)_Y \otimes U(1)' \otimes \mathbb{Z}_2$
symmetry. Here $i = 1$, 2, 3 and $k = 1$, 2.}
\label{Themodel}
\end{table}

As can be seen in Table~\ref{Themodel}, the scalar field $\eta $ has a non-trivial $\mathbb{Z}_2$ charge and corresponds to the inert $SU(2)_L$ scalar doublet, implying that the lightest electrically neutral CP-even and CP-odd components can be viable scalar DM candidates. Furthermore, having non-trivial charges under the preserved $\mathbb{Z}_2$ discrete symmetry, the scalar singlets $\varphi$, $\rho$, and $\zeta$ do not acquire a vacuum expectation value (VEV), thus possibly providing scalar DM candidates with the lightest CP-even or odd components. In addition, to this end, the lightest of the two states $N_{R_{k}}$ can be a fermionic DM candidate. Furthermore, the global symmetry $U(1)'$ is spontaneously broken by the VEV (at the TeV scale) of the gauge-singlet scalar field $\sigma$. 
With this setup, as will be shown below, the scalar fields $\eta$, $\rho$, $\zeta$, and $\varphi$ play a key role in the implementation of a three-loop radiative seesaw mechanism for the generation of light and active neutrino masses. Furthermore, neutral fermions $N_{R_{k}}$, when they are of  Majorana nature,  mediate the radiative seesaw mentioned above. 
Due to the $\mathbb{Z}_2$ symmetry, the RH neutrinos do not mix with active neutrinos, preventing the latter from acquiring tree-level masses. 
Finally, the $U(1)' \otimes \mathbb{Z}_2$ symmetry is crucial to prevent one-loop and two-loop masses for light-active neutrinos.

Given the symmetries and particle spectrum described, the allowed charged-fermion and neutrino Yukawa interactions are, respectively,
\begin{align}
    -\mathcal{L}_Y &=y_{u \phi}^{ij}\, \bar{q}_{iL} \widetilde{\phi} u_{jR} + y_{d \phi}^{ij}\, \bar{q}_{iL} \phi d_{jR} + y_{l \phi}^{ij}\, \bar{\ell}_{iL} \phi \ell_{R_j} + \mathrm{H.c.}\\
    -\mathcal{L}_Y^{(\nu)} &= y_\eta^{ik}\, \bar{\ell}_{iL} \widetilde{\eta} N_{R_{k}} + M_{N_{R}}^{kr}\, \bar{N}_{R_{k}} N_{R_{r}}^C + \mathrm{H.c.}
\label{eq:lagrangian2}\ ,\end{align}
where the superscript $C$ stands for charge conjugation $(\Psi^C \equiv C\, \overline{\Psi}^T)$, $\widetilde{\phi} = i\, \tau_2\, \phi^*$, $\widetilde{\eta} = i\, \tau_2\, \eta^*$, and the summation over repeated indices must be understood, with $i,\, j = 1,\, 2,\, 3$ and $k,\, r = 1,\, 2$.
\subsection{The scalar potential}
The most general scalar potential invariant under the symmetries of the
model is
\begin{align}
    V& =-\mu _{\phi }^{2}(\phi ^{\dagger }\phi )-\mu _{\sigma }^{2}(\sigma^{\ast }\sigma )+\mu _{\eta }^{2}(\eta ^{\dagger }\eta )+\mu _{\varphi}^{2}(\varphi ^{\ast }\varphi )+\mu _{\rho }^{2}(\rho ^{\ast }\rho )+\mu_{\zeta }^{2}(\zeta ^{\ast }\zeta )+\widetilde{\mu }_{\zeta }^{2}\left(\zeta ^{2}+\mathrm{H.c.}\right)   \notag \\
    & \quad +\lambda _{1}(\phi ^{\dagger }\phi )^{2}+\lambda _{2}(\sigma ^{\ast}\sigma )^{2}+\lambda _{3}(\phi ^{\dagger }\phi )(\sigma ^{\ast }\sigma)+\lambda _{4}(\eta ^{\dagger }\eta )^{2}+\lambda _{5}(\varphi ^{\ast}\varphi )^{2}+\lambda _{6}(\rho ^{\ast }\rho )^{2}  \notag \\
    & \quad +\lambda _{7}(\zeta ^{\ast }\zeta )^{2}+\left( \kappa _{1}\zeta ^{4}+ \mathrm{H.c.}\right) +\left( \kappa _{2}\zeta ^{2}+\mathrm{H.c.}\right)(\zeta ^{\ast }\zeta )+\lambda _{8}(\eta ^{\dagger }\eta )(\varphi ^{\ast}\varphi )+\lambda _{9}(\eta ^{\dagger }\eta )(\rho ^{\ast }\rho )  \notag \\
    & \quad +\lambda _{10}(\eta ^{\dagger }\eta )\zeta ^{2}+\left( \kappa_{3}\zeta ^{2}+\mathrm{H.c.}\right) (\eta ^{\dagger }\eta )+\lambda_{11}(\varphi ^{\ast }\varphi )(\rho ^{\ast }\rho )+\lambda _{12}(\varphi^{\ast }\varphi )(\zeta ^{\ast }\zeta )  \notag \\
    & \quad +\left( \kappa _{4}\zeta ^{2}+\mathrm{H.c.}\right) (\varphi ^{\ast}\varphi )+\lambda _{13}(\rho ^{\ast }\rho )(\zeta ^{\ast }\zeta )+\left(\kappa _{5}\zeta ^{2}+\mathrm{H.c.}\right) (\rho ^{\ast }\rho )+\lambda_{14}\left( \varphi \rho ^{3}+\mathrm{H.c.}\right)   \notag \\
    & \quad +\lambda _{15}\left( \rho \zeta \sigma ^{2}+\mathrm{H.c.}\right)+\lambda _{16}(\phi ^{\dagger }\phi )(\eta ^{\dagger }\eta )+\lambda_{17}(\phi ^{\dagger }\eta )(\eta ^{\dagger }\phi )+\lambda _{18}(\phi^{\dagger }\phi )(\varphi ^{\ast }\varphi )  \notag \\
    & \quad +\lambda _{19}(\phi ^{\dagger }\phi )(\rho ^{\ast }\rho )+\lambda_{20}(\phi ^{\dagger }\phi )(\zeta ^{\ast }\zeta )+\left( \kappa _{6}\zeta^{2}+\mathrm{H.c.}\right) (\phi ^{\dagger }\phi )+\lambda _{21}(\sigma^{\ast }\sigma )(\eta ^{\dagger }\eta )  \notag \\
    & _{\quad }+\lambda _{22}(\sigma ^{\ast }\sigma )(\varphi ^{\ast }\varphi)+\lambda _{23}(\sigma ^{\ast }\sigma )(\rho ^{\ast }\rho )+\lambda_{24}(\sigma ^{\ast }\sigma )(\zeta ^{\ast }\zeta )+\left( \kappa _{7}\zeta^{2}+\mathrm{H.c.}\right) \left( \sigma ^{\ast }\sigma \right)   \notag \\
    & _{\quad }+A\left[ (\eta ^{\dagger }\phi )\varphi +\mathrm{H.c.}\right] ,
    \label{eq:sacalar-potential}  
\end{align}
where the coefficients $\mu_\phi$, $\mu_\sigma$, $\mu_\eta$, $\mu_\varphi$, $\mu_\rho$, $\mu_\zeta$, $\widetilde{\mu }_{\zeta }$ and the trilinear coupling $A$ are dimensionful parameters, while the remaining quartic couplings are dimensionless. The scalar fields of the model can be expanded as 
\begin{align} 
\phi & =\left( 
\begin{array}{c}
\phi ^{+} \\ 
\frac{1}{\sqrt{2}}\left( v+\phi _{R}^{0}+i\,\phi _{I}^{0}\right) 
\end{array}%
\right) ,\hspace{1cm}\eta =\left( 
\begin{array}{c}
\eta ^{+} \\ 
\frac{1}{\sqrt{2}}\left(\eta _{R}^{0}+i\,\eta _{I}^{0}\right) 
\end{array}
\right) ,  \notag \\
\sigma & =\frac{1}{\sqrt{2}}\left( v_{\sigma }+\sigma _{R}+i\,\sigma
_{I}\right) ,\hspace{1cm}\rho =\frac{1}{\sqrt{2}}\left( \rho _{R}+i\,\rho
_{I}\right) ,\hspace{1cm}  \notag \\
\varphi & =\frac{1}{\sqrt{2}}\left( \varphi _{R}+i\,\varphi _{I}\right) ,%
\hspace{1cm}\zeta =\frac{1}{\sqrt{2}}\left( \zeta _{R}+i\,\zeta _{I}\right) .
\end{align}

The scalar potential minimization conditions yield the following relations 
\begin{align}
    \mu _{\phi }^{2}& =\lambda _{1}\,v^{2}+\frac{1}{2}\,\lambda _{3}\,v_{\sigma}^{2}\,,  \notag \\
    \mu _{\sigma }^{2}& =\frac{\lambda _{3}\,v^{2}}{2}+\lambda _{2}\,v_{\sigma}^{2}\,.
\end{align}
From the analysis of the scalar potential, it follows that $\phi _{I}^{0}$
and $\sigma _{I}$ are massless CP-odd scalar fields corresponding to the SM
neutral Goldstone boson and to the majoron, respectively.
Regarding the remaining CP-odd scalar fields, the only mixings allowed by the symmetries of the model at tree level are $\eta _{I}-\varphi _{I}$ and $\rho_{I}-\zeta _{I}$. The corresponding CP-odd squared mass matrices are given, in the basis $\left( \eta _{I},\,\varphi_{I}\right) $ and $\left( \rho _{I},\,\zeta _{I}\right) $ by
\begin{align}
M_{A}^{2} &=\left( 
\begin{array}{cc}
\frac{1}{2}\left[ 2\,\mu _{\eta }^{2}+\left( \lambda _{16}+\lambda
_{17}\right) \,v^{2}+\lambda _{21}\,v_{\sigma }^{2}\right]  & \frac{A\,v}{%
\sqrt{2}} \\ 
\frac{A\,v}{\sqrt{2}} & \frac{1}{2}\left( 2\,\mu _{\varphi }^{2}+\lambda
_{18}\,v^{2}+\lambda _{22}\,v_{\sigma }^{2}\right) 
\end{array}%
\right), \\
M_{P}^{2} &=\left( 
\begin{array}{cc}
\frac{1}{2}\left( 2\,\mu _{\rho }^{2}+\lambda _{19}\,v^{2}+\lambda
_{23}\,v_{\sigma }^{2}\right)  & -\frac{1}{2}\lambda _{15}v_{\sigma }^{2} \\ 
-\frac{1}{2}\lambda _{15}v_{\sigma }^{2} & \frac{1}{2}\left[ 2\,\mu _{\zeta
}^{2}-4\widetilde{\mu }_{\zeta }^{2}+\left( \lambda _{20}-2\kappa
_{6}\right) \,v^{2}+\left( \lambda _{24}-2\kappa _{7}\right) \,v_{\sigma
}^{2}\right] 
\end{array}
\right).
\end{align}
On the other hand, in the CP-even scalar sector, the only mixings allowed by
the symmetries of the model at tree level are $\phi _{R}-\sigma _{R}$, $\eta _{R}-\varphi
_{R}$, and $\rho _{R}-\zeta_{R} $. The corresponding squared CP-even mass
matrices are given, in the basis $\left( \phi _{R},\,\sigma _{R}\right) $, $%
\left( \eta _{R},\,\varphi _{R}\right) $, $\left( \rho _{R},\,\zeta_{R} \right)$, by 
\begin{align}
M_{H}^{2}& =\left( 
\begin{array}{cc}
2\,\lambda _{1}\,v^{2} & \lambda _{3}\,v\,v_{\sigma } \\ 
\lambda _{3}\,v\,v_{\sigma } & 2\,\lambda _{2}\,v_{\sigma }^{2}%
\end{array}%
\right) ,  \notag \\
M_{\Phi }^{2}& =M_{A}^{2}=\left( 
\begin{array}{cc}
\frac{1}{2}\left[ 2\,\mu _{\eta }^{2}+\left( \lambda _{16}+\lambda
_{17}\right) \,v^{2}+\lambda _{21}\,v_{\sigma }^{2}\right]  & \frac{A\,v}{%
\sqrt{2}} \\ 
\frac{A\,v}{\sqrt{2}} & \frac{1}{2}\left( 2\,\mu _{\varphi }^{2}+\lambda
_{18}\,v^{2}+\lambda _{22}\,v_{\sigma }^{2}\right) 
\end{array}%
\right) ,  \notag \\
M_{\Xi }^{2}& =\left( 
\begin{array}{cc}
\frac{1}{2}\left( 2\,\mu _{\rho }^{2}+\lambda _{19}\,v^{2}+\lambda
_{23}\,v_{\sigma }^{2}\right)  & \frac{1}{2}\lambda _{15}v_{\sigma }^{2} \\ 
\frac{1}{2}\lambda _{15}v_{\sigma }^{2} & \frac{1}{2}\left[ 2\,\mu _{\zeta
}^{2}+4\widetilde{\mu }_{\zeta }^{2}+\left( \lambda _{20}+2\kappa
_{6}\right) \,v^{2}+\left( \lambda _{24}+2\kappa _{7}\right) \,v_{\sigma
}^{2}\right] 
\end{array}%
\right). \label{eq:massmatrixCPodd}
\end{align}

The squared mass matrices for the neutral CP even $M_{H}^{2}$, $M_{\Phi }^{2}$, $M_{\Xi }^{2}$ and neutral CP odd $M_{A}^{2}$, $M_{P}^{2}$ scalars can be diagonalized by the following transformations 
\begin{equation}
R_{S}^{T}\,M_{S}^{2}\,R_{S} =\left( 
\begin{array}{cc}
\frac{X_{S}+Y_{S}}{2}+\frac{1}{2}\sqrt{\left( X_{S}-Y_{S}\right)
^{2}+4\,Z_{S}^{2}} & 0 \\ 
0 & \frac{X_{S}+Y_{S}}{2}-\frac{1}{2}\sqrt{\left( X_{S}-Y_{S}\right)
^{2}+4\,Z_{S}^{2}}%
\end{array}%
\right),
\end{equation}
where
\begin{align}
R_S& =\left( 
\begin{array}{cc}
\cos \theta_S & -\sin \theta _{S} \\ 
\sin \theta_S & \cos \theta_S%
\end{array}%
\right) ,\hspace{1.2cm}\theta _{\Phi }=\theta _{A},\hspace{1.2cm}\text{with}%
\quad S=H,\Phi ,\,\Xi ,A,P,\,  \notag \\
X_{S}& =\left( M_{S}^{2}\right) _{11}\,,\hspace{1.2cm}Y_{S}=\left(
M_{S}^{2}\right) _{22}\,,\hspace{1.2cm}Z_{S}=\left( M_{S}^{2}\right) _{12}\,,
\notag \\
\tan 2\theta _{S}& =\frac{2\,Z_{S}}{X_{S}-Y_{S}}=\frac{2\left(
M_{S}^{2}\right) _{12}}{\left( M_{S}^{2}\right) _{22}-\left(
M_{S}^{2}\right) _{11}}\,.
\end{align}%
Finally, $\phi ^{\pm }$ are the SM charged Goldstone bosons, whereas $\eta
^{\pm }$ are the physical electrically charged scalars with squared mass 
\begin{equation}
    m_{\eta ^{\pm }}^{2}=\frac{1}{2}\left( 2\,\mu _{\eta }^{2}+\lambda_{16}\,v^{2}+\lambda _{21}\,v_{\sigma }^{2}\right).
\end{equation}

\subsection{Accidental symmetries and smallness of neutrino masses}
As will be shown in the following, light-active neutrino masses vanish when $\theta_\Phi \to 0$, which corresponds to the absence of the trilinear scalar interaction $A \left[(\eta^\dagger \phi) \varphi + \mathrm{H.c.}\right]$ in Eq.~(\ref{eq:sacalar-potential}). In this limit, the Yukawa interactions and the scalar potential have an accidental global symmetry $U(1)_X$,  under which the inert $SU(2)_L$ scalar doublet $\eta$ has charge $-1$ and the charges of the left and right leptonic fields are equal to $1$, while the remaining fields are neutral. Therefore, this global $U(1)_X$ does not undergo spontaneous breaking, leaving the particle spectrum free of an extra Goldstone boson.
However, the trilinear scalar coupling $A$ breaks the $U_X(1)$ softly. 
As a result, the small masses of light neutrinos are protected in our model by this accidental symmetry and therefore are technically natural.

\section{Neutrino mass matrix} \label{Sec:Neutrino}
\begin{figure}[t!]
    \begin{center}
        \includegraphics[width=0.55\textwidth]{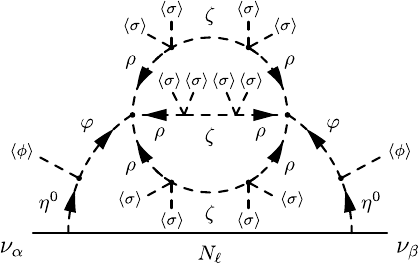} 
    \end{center}
    \caption{Scotogenic loop for light active neutrino masses where $\ell=1$, 2 and $\protect\alpha,\, \protect\beta = e,\, \protect\mu,\, \protect\tau$.}
    \label{Neutrinodiagram}
\end{figure}
The three-loop contribution to the ($3\times 3$ light and active) neutrino mass matrix is shown in Fig.~\ref{Neutrinodiagram} and is formulated as
\begin{equation} \label{Mnu}
    (m _\nu) _{\alpha \beta} = \sum _\ell\, \left(y_{\alpha \ell}\, \Lambda _\ell\, y_{\beta \ell}\right),
\end{equation}
where
\begin{align} \label{eq:lambda}
    \Lambda _\ell & = \frac{9}{4}\lambda _{14} ^2 \frac{\cos ^2 \theta _\Phi \sin ^2 \theta _\Phi}{(4 \pi) ^6} \Delta m _\Phi ^4 \sum _{i, j, k} M _\ell \, a _i\, a _j\, a _k  \nonumber\\
    & \times \int _0 ^1 dx\,dy\, \delta(x + y - 1) \int _0 ^1 dz\, dw\, \delta(z + w - 1) \int _0 ^1 ds\, dt\, du\, dv\, \delta(s + t + u + v - 1) \nonumber\\
    & \times \frac{s\, t}{x\, y\, z\, w} \frac{1}{M _{i j k \ell} ^8} \left(y\, m _j ^2 - \frac{x ^2}{y} m _k ^2 \right) \left(\frac{z ^2}{w} m _i ^2 - \frac{w}{x} m _j ^2 - \frac{w}{y} m _k ^2 \right) .
\end{align}
In the above expression, $\Delta m_\Phi^2 \equiv m_{\Phi_1}^2 - m_{\Phi_2}^2$ is the squared mass difference between the mass eigenvalues of $\eta^0$ and $\varphi$ given by the matrices $M_{\Phi}^2$ and $M_{A}^2$, defined in Eq.~\eqref{eq:massmatrixCPodd}. 
The indices $i,\, j,\, k=1,\, 2$ represent the mass eigenstates for the CP-even  states in the matrix $M_{\Xi}^2$, while the indices $i,\, j,\, k=3,\, 4$ are for the CP-odd states in the matrix $M_{P}^2$.
The index $\ell=1,\, 2$ labels the RH neutrinos. 
Furthermore, the coefficients $a_i$ are given by 
\begin{equation}
    a_{i} = 
    \begin{dcases}
        +\cos^2\theta _\Xi  & \text{for } i = 1\,, \\ 
        +\sin^2\theta _\Xi  & \text{for } i = 2\,,\\
        -\cos^2\theta_\Xi ^\prime                 & \text{for } i=3\,,\\
        -\sin^2\theta_\Xi ^\prime                 & \text{for } i=4\,.
    \end{dcases}
\end{equation}
Finally, the mass parameter $M_{ijk\ell}^2$ is defined by 
\begin{equation}
    M_{ijk\ell}^2 \equiv s\, m_{\Phi_1}^2 + t\, m_{\Phi_2}^2 + \frac{u}{w} m_i^2 + \frac{u}{x\, z} m_j^2 + \frac{u}{y\, z} m_k^2 + v\, M_\ell^2,
\end{equation}
and the mixing angles $\theta_\Phi$ for the complex scalars, $\theta_\Xi$ for the CP-even and $\theta_\Xi ^\prime$ for the CP-odd real scalars, are defined by the matrices
\begin{align}
    \begin{pmatrix}
        \eta^{0} \\ 
        \varphi
    \end{pmatrix}
    &=
    \begin{pmatrix}
        \cos \theta_\Phi & \sin \theta_\Phi \\ 
        -\sin \theta_\Phi & \cos \theta_\Phi
    \end{pmatrix}
    \begin{pmatrix}
        \Phi_{1} \\ 
        \Phi_{2}
    \end{pmatrix},\\
    \begin{pmatrix}
        \rho_R \\ 
        \zeta_R
    \end{pmatrix}
    &=
    \begin{pmatrix}
        \cos \theta_{\Xi } & \sin \theta_{\Xi } \\ 
        -\sin \theta_{\Xi } & \cos \theta_{\Xi }
    \end{pmatrix}
    \begin{pmatrix}
        \Xi_{1} \\ 
        \Xi_{2}
    \end{pmatrix},\\
   \begin{pmatrix}
        \rho_I \\ 
        \zeta_I
    \end{pmatrix}
    &=
    \begin{pmatrix}
        \cos \theta_\Xi' & \sin \theta_{\Xi }^\prime \\ 
        -\sin \theta_\Xi' & \cos \theta_{\Xi }^\prime
    \end{pmatrix}
    \begin{pmatrix}
        \Xi_{3} \\ 
        \Xi_{4}
    \end{pmatrix}.
\end{align}

It is interesting to note that in the original scotogenic model~\cite{Ma:2006km}, the scalar quartic coupling $\left(\phi^\dag\eta\right)^2$ arises at tree level. However, in the present scenario, it is induced at the two-loop level, as shown in Fig.~\ref{Neutrinodiagram}, being therefore naturally suppressed.
This coupling should be small enough, as required by the experimental constraints~\cite{Kubo:2006yx, Suematsu:2009ww, Schmidt:2012yg}. 

The neutrino mass matrix is diagonalized by the PMNS matrix (Pontecorvo-Maki-Nakagawa-Sakata) $U _{\text{PMNS}}$ defined by
\begin{equation}
    U _{\text{PMNS}} ^T \, m _\nu \, U _{\text{PMNS}} = \text{diag} ( m _1, m _2, m _3 ) .
\end{equation}
Notice that since we only have two RH neutrinos, the neutrino mass matrix thus has two non-vanishing eigenvalues, so that one has a massless neutrino in the spectrum 
($\nu _1$ in the normal ordering (NO) or $\nu _3$ in the inverted ordering (IO)).
The matrix $U _{\text{PMNS}}$ is parameterized as
\begin{equation}
\label{eq:PMNS}
U _{\text{PMNS}} = \begin{pmatrix}  c_{12}c_{13} & s_{12}c_{13}  & s_{13}e^{i\delta}  \\
-s_{12}c_{23}-c_{12}s_{23}s_{13}e^{-i\delta}  & 
c_{12}c_{23}-s_{12}s_{23}s_{13}e^{-i\delta}  & s_{23}c_{13}  \\
s_{12}s_{23}-c_{12}c_{23}s_{13}e^{-i\delta}  & 
-c_{12}s_{23}-s_{12}c_{23}s_{13}e^{-i\delta}  & c_{23}c_{13} \end{pmatrix} 
\begin{pmatrix} e^{i\varphi_1/2} & 0 & 0 \\
0 & e^{i\varphi_2/2}  & 0 \\
0 & 0 & 1 \end{pmatrix} ,
\end{equation}
where $c_{ij} = \cos \theta_{ij}$, $s_{ij} = \sin \theta_{ij}$, $\delta$ is the Dirac phase and $\varphi_1$, $\varphi_2$ are the Majorana phases. In our numerical study, we use the best-fit values of the NuFIT 5.0 global analysis~\cite{Esteban:2020cvm}
\begin{align}
    \sin ^2 \theta _{12} = 0.304, \quad \sin^2 \theta_{13} = 0.02219, \quad \sin^2\theta _{23} = 0.573, \nonumber\\
    \Delta m _{21} ^2 = 7.42 \times 10 ^{-5} \, \text{eV} ^2 , \quad \Delta m _{3l} ^2 = 2.517 \times 10 ^{-3} \, \text{eV} ^2 , \label{eq:nu_oscil_parameters}
\end{align}
where $\Delta m _{3l} ^2 \equiv \Delta m _{31} ^2 > 0$ for NO and $\Delta m _{3l} ^2 \equiv \Delta m _{32} ^2 < 0$ for IO.

It is convenient to express the Yukawa couplings in terms of the oscillation parameters using an adapted Casas-Ibarra parameterization~\cite{Casas:2001sr, Cordero-Carrion:2019qtu, Restrepo:2019ilz}
\begin{equation} \label{eq:adaptedCasasIbarra}
    y _{\eta} = \sqrt{\Lambda} ^{-1} \, R \, \begin{pmatrix} 0 & \sqrt{m _2} & 0 \\ 0 & 0 & \sqrt{m _3} \end{pmatrix} P \, U _{\text{PMNS}} ^\dagger\,,
\end{equation}
where $\Lambda$ is defined as $\Lambda = \text{diag} (\Lambda _1, \Lambda _2)$, with $\Lambda _\ell$ given by Eq.~\eqref{eq:lambda}. $R$ is a general orthogonal $2 \times 2$ complex matrix satisfying $R ^T R = R\, R ^T = \mathbb{I} _2$, and the matrix $P$ depends on the ordering of the neutrino spectrum, defined as $P = \mathbb{I} _3$ for NO and $P = P _{13}$ for IO, with 
\begin{equation}
    P _{13} = \begin{pmatrix} 0 & 0 & 1 \\ 0 & 1 & 0 \\ 1 & 0 & 0 \end{pmatrix} .
\end{equation}
In the case of IO, one should also replace $m _3 \to m _1$. Note that the peculiar form of this parameterization stems from the fact that we have only two RH neutrinos. This fact is also reflected in the number of degrees of freedom of the matrix $R$, which can be parameterized by a single complex angle, in contrast to the three complex angles in the three RH neutrino case. Although the $R$ matrix generally contains two free parameters, for simplicity, we will assume that it depends on only one real angle. Furthermore, in what follows, we assume that the Dirac and Majorana complex phases vanish, so that the Yukawa couplings, as given by Eq.~\eqref{eq:adaptedCasasIbarra}, are real.

\subsection{Effective mass for neutrinoless double beta decay}
In our model, only two RH neutrinos mediate the loop diagram; therefore, one eigenvalue will be massless, and the obtained spectrum for light neutrino masses can accommodate both orderings of the spectrum (that is, inverted and normal).
Furthermore, the first entry of the mass matrix obtained in Eq.~\eqref{Mnu} is related to the effective mass in neutrinoless double beta decay ($0\nu 2\beta$). 
In this case, its tree-level decay amplitude is proportional to 
\begin{equation}\label{Gamma0nubb}
    \sum G^2_F\, {\bf U}^2_{ei} \, \gamma_\mu \, P_R \, \frac{\slash \hspace*{-2.5mm}p + m_i}{p^2 - m_i^2} \, \gamma_\nu \, P_L \,\simeq \, \sum G^2_F\, {\bf U}^2_{ei} \, \frac{m_i}{p^2} \,\gamma_\mu \, P_R \,\gamma_\nu\,,
\end{equation}
where $G_F$ is the Fermi constant, 
$m_i$ the neutrino (physical) mass, and $p$ is the virtual momentum of the neutrino such that $p^2 \simeq - (125~\mbox{MeV})^2$ (the value corresponding to an average of the virtual momenta in different decaying nuclei).
We recall that in our setup, the two additional neutrinos $N_{R_{1,2}}$ do not contribute to the decay amplitude due to the $\mathbb{Z}_2$ symmetry; see Eq.~\eqref{eq:lagrangian2}.
Finally, the $0\nu 2 \beta$ decay width is only proportional to the effective mass and not to the individual masses,\footnote{It is worth emphasizing that given our setup, there cannot be any further contributions to this effective mass.} due to the contribution of light-active neutrinos
\begin{equation} \label{mee}
    m_{ee}\,=\,\left|\sum_{i=1}^3 {\bf U}^2_{ei}\,m_i \right|\,.
\end{equation}

Plugging the neutrino oscillation parameters given in Eq.~\eqref{eq:nu_oscil_parameters} into the previous expression, one can easily see that if we neglect the complex phases, the effective mass has a fixed value for each neutrino mass ordering, namely, $m _{ee} ^{\text{NO}} = 3.67$~meV (for NO) and $m _{ee} ^{\text{IO}} = 48.36$~meV (for IO). Both of these values satisfy the KamLAND-ZEN upper limit $m _{ee} \lesssim 61$~meV~\cite{KamLAND-Zen:2016pfg}. 

While in the NO case $m _{ee}$ is too small to be detected in future $0 \nu 2 \beta$ experiments, for the IO $m _{ee}$ is within the reach of experiments such as EXO~\cite{Tosi:2014zza}, LEGEND~\cite{LEGEND:2017cdu}, CUPID~\cite{CUPID-0:2018rcs} and NEXT~\cite{NEXT:2013wsz}. Therefore, a positive $0 \nu 2 \beta$ signal in the near future would mean to our model that the neutrino mass ordering must be the inverted one. This picture does not change if we allow for general complex phases. Note that the values for $m _{ee} ^{\text{NO}}$ and $m _{ee} ^{\text{IO}}$ reported above correspond to the {\it maximum} possible values for $m _{ee}$, as the effective mass is suppressed by cancelation between the different phases when on. Therefore, when we continuously vary the complex phases, we would see a distribution of $m _{ee}$ values with $m _{ee} ^{\text{NO}}$ and $m _{ee} ^{\text{IO}}$ being the maximum in each ordering scenario. Therefore, there would still be a region of parameter space that would be within the experimental reach in the case of an IO. On the other hand, $0 \nu 2 \beta$ would remain beyond the reach of next-generation experiments in the NO case.

\section{Phenomenological implications} \label{sec:cLFVobs}

In this section, the impact of the model on electroweak precision measurements and charged-lepton flavor violation is studied. We first address the impact of our model on the oblique parameters and the $W$ boson mass, and then consider some cLFV observables in the $\mu-e$ sector, from which we have stringent bounds.
\subsection{Oblique parameters and $W$ boson mass}
The extra scalars in our model, in particular those of the inert scalar doublet, affect the oblique corrections of the SM, which are parameterized in terms of the well-known quantities $T$, $S$ and $U$, defined as~\cite{Altarelli:1990zd, Peskin:1991sw, Barbieri:2004qk} 
\begin{align}
    T &= \left. \frac{\Pi_{33}\left(q^2\right) - \Pi_{11}\left(q^2\right)}{\alpha_\text{EM}(M_Z)\, M_W^2} \right|_{q^2=0},\\
    S &= \left.\frac{2\, \sin 2 \theta_W}{\alpha_\text{EM}(M_Z)} \frac{d\Pi_{30}\left(q^2\right)}{dq^2} \right|_{q^2=0}, \\
    U &= \left.\frac{4 \sin^2\theta_W}{\alpha_\text{EM}(M_Z)} \left(\frac{d\Pi_{33}\left(q^2\right)}{dq^2} - \frac{d\Pi_{11}\left(q^2\right) }{dq^2}\right)\right|_{q^2=0},
\end{align}
where $\Pi_{11}\left(0\right) $, $\Pi_{33}\left(0\right) $, and $\Pi_{30}\left(0\right)$ are the vacuum polarization amplitudes with the $\{W_\mu^1\, ,W_\mu^1\}$, $\{W_{\mu}^3,\, W_\mu^3\}$ and $\{W_\mu^3,\, B_\mu\}$ external gauge bosons, respectively. We note that in the definition of the parameters $S$, $T$, and $U$, the new physics scale is assumed to be heavy compared to the masses of the electroweak gauge bosons $W$ and $Z$, $M_W$ and $M_Z$, respectively. Because their values are measured in high-precision experiments, they act as constraints on our model. Furthermore, the new measurement of the $W$ gauge boson mass by the CDF collaboration~\cite{CDF:2022hxs}, can be interpreted as an indication of non-trivial $S$, $T$, and $U$ values. Namely, non-SM particles provide radiative corrections to the $W$ boson mass through their contribution to the oblique parameters
\begin{equation}
    M_W^2 = \left(M_W^2\right)_\text{SM} + \frac{\alpha_\text{EM}\left(M_Z\right) \cos^2\theta_W\, M_Z^2}{\cos^2\theta_W - \sin^2\theta_W} \left[-\frac{S}{2} + \cos^2\theta_W\, T + \frac{\cos^2\theta_W - \sin^2\theta_W}{4\, \sin^2\theta_W}\, U\right] ,
\end{equation}
where, as pointed out in Ref.~\cite{Peskin:1991sw}, $\left(M_W^2\right)_\text{SM}$ is the mass of the $W$ gauge boson determined in the SM model with the highest possible accuracy. Therefore, the observed shift in the $W$ boson mass can be explained by new physics that contributes to the oblique parameters. In this section, we calculate the one-loop contributions to $S$, $T$, and $U$ in our model and find the parameter space in which the CDF result can be accommodated.

The one-loop contributions to the oblique parameters arising from the inert scalar exchange are
\begin{align}
\label{eq:stu}
    T &\simeq \frac{1}{16 \pi^2\, v^2\, \alpha_\text{EM}(M_Z)} \left[\sum_{i=1}^2 \sum_{j=1}^2 \left(\left(R_\Phi\right)_{1i}\right)^2 \left(\left(R_A\right)_{1j}\right)^2\, F\left(m_{\Phi_i^0}^2,\, m_{A_j^0}^2\right) + m_{\eta^\pm}^2\right.  \notag\\
    &\qquad -\left. \sum_{i=1}^2 \left(\left(R_\Phi\right)_{1i}\right)^2\, F\left(m_{\Phi_i^0}^2,\, m_{\eta^\pm}^2\right) - \sum_{i=1}^2 \left(\left(R_A\right)_{1i}\right)^2\, F\left(m_{A_i^0}^2,\, m_{\eta^\pm}^2\right) \right],\\
\label{eq:stu2}
    S &\simeq \sum_{i=1}^2 \sum_{j=1}^2 \frac{\left(\left(R_\Phi\right)_{1i}\right)^2 \left(\left(R_A\right)_{1j}\right)^2}{12 \pi}\, K\left(m_{\Phi_i^0}^2,\, m_{A_j^0}^2,\, m_{\eta^\pm}^2\right),\\
\label{eq:stu3}
    U &\simeq -S + \sum_{i=1}^2 \left[\left(\left(R_A\right)_{1i}\right)^2\, K_2\left(m_{A_j^0}^2,\, m_{\eta^\pm}^2\right) + \left(\left(R_\Phi\right)_{1i}\right)^2\, K_2\left(m_{\Phi_i^0}^2,\, m_{\eta^\pm}^2\right)\right],
\end{align}
where we introduce the functions~\cite{CarcamoHernandez:2015smi} 
\begin{align}
    F\left(m_1^2,\, m_2^2\right) &= \frac{m_1^2\, m_2^2}{m_1^2 - m_2^2}\, \ln\left(\frac{m_1^2}{m_2^2}\right),\\
    K\left(m_1^2,\, m_2^2,\, m_3^2\right) &= \frac{1}{\left(m_2^2-m_1^2\right)^3} \left[m_1^4 \left(3\, m_2^2 - m_1^2\right)\, \ln\left(\frac{m_1^2}{m_3^2}\right) - m_2^4 \left(3\, m_1^2 - m_2^2\right) \ln\left(\frac{m_2^2}{m_3^2}\right)\right.  \notag \\
    &\qquad -\left.\frac16 \left[27\, m_1^2 m_2^2 \left(m_1^2 - m_2^2\right) + 5 \left(m_2^6 - m_1^6\right)\right]\right],
\end{align}
with the following properties 
\begin{align}
    & \lim_{m_2 \to m_1} F\left(m_1^2,\, m_2^2\right) = m_1^2\,,\\
    & \lim_{m_1 \to m_3} K(m_1^2,\, m_2^2,\, m_3^2) = K_2(m_2^2,\, m_3^2)\,,\\
    & \lim_{m_1 \to m_2} K(m_1^2,\, m_2^2,\, m_3^2) = K_1(m_2^2,\, m_3^2) = \ln\left(\frac{m_2^2}{m_3^2}\right),\\
    & \lim_{m_2 \to m_3} K(m_1^2,\, m_2^2,\, m_3^2) = K_2(m_1^2,\, m_3^2) \nonumber\\
    &\qquad\qquad\qquad = \frac{- 5 m_1^6 + 27 m_1^4 m_3^2 - 27 m_1^2 m_3^4 + 6 \left(m_1^6 - 3 m_1^4 m_3^2\right) \ln\left(\frac{m_1^2}{m_3^2}\right) + 5 m_3^6}{6 \left(m_1^2 - m_3^2\right)^3}\,.
\end{align}
Here $R_\Phi$ and $R_A$ are the rotation matrices diagonalizing the
squared mass matrices $M_\Phi^2$ and $M_A^2$, respectively,
according to
\begin{equation}
    R_{\Phi,\, A}^T\, M_{\Phi,\, A}^2\, R_{\Phi,\,A} = \left(M_{\Phi,\, A}^2\right)_\text{diag}.
\end{equation}
Note that, since the matrices $M _\Phi ^2$ and $M _A ^2$ are identical (cf. Eq.~\eqref{eq:massmatrixCPodd}), it follows that $m _{A _i ^0} = m _{\Phi _i ^0}$ and $(R _A) _{ij} = (R _\Phi) _{ij}$, so that $S$, $T$, $U$ can be chosen to be functions of $m _{\eta ^\pm}$, $m _{\Phi _i ^0}$ and $\theta _\Phi$. Note also that, in the degenerate limit $m _{\Phi _1 ^0} = m _{\Phi _2 ^0} = m _{\eta ^\pm}$, all oblique parameters vanish. Therefore, $\Phi _1 ^0$, $\Phi_2^0$, and $\eta ^\pm$ must be non-degenerate to give rise to non-vanishing contributions.

\begin{figure}[t!]
    \centering
    \includegraphics[width=0.48\linewidth]{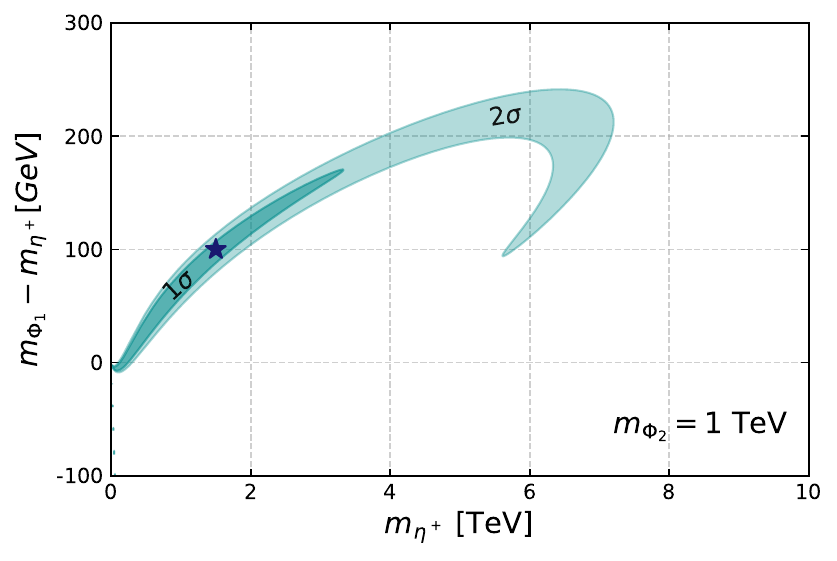}\quad
    \includegraphics[width=0.48\linewidth]{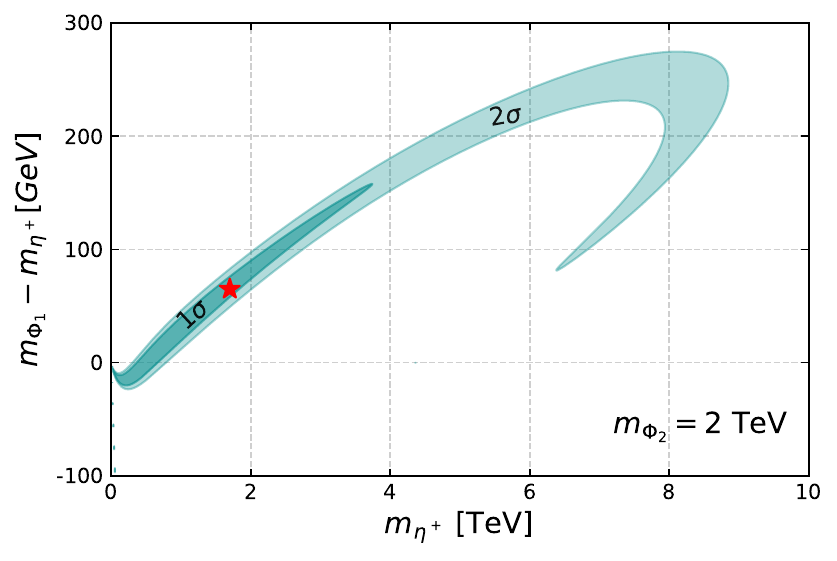}
    \caption{The $1$-$\protect\sigma$ and $2$-$\protect\sigma$ regions in the $m_{\Phi_1} - m_{\protect\eta^+}$ versus $m_{\protect\eta^+}$ plane allowed by the fit of the oblique $S$, $T$ and $U$ parameters including the CDF measurement of the $W$ mass. In the left (right) panel the mass of the neutral scalar $\Phi_2$ is $m_{\Phi_2} = 1$~TeV ($m_{\Phi_2} = 2$~TeV). In both cases, the mixing angle is fixed at $\protect\theta_\Phi = 0.2$.}
    \label{stuCDFplot}
\end{figure}

\begin{table}[t]
  \centering
  \begin{tabular}{|c|c|c c c|}
  \hline
    & \textbf{Parameter Value}  & \multicolumn{3}{c|}{\textbf{Correlation}} \\
    &                 & $S$    & $T$    & $U$ \\
    \hline
$S$ & $0.06 \pm 0.10$ & 1.00 & 0.90 & -0.59 \\
$T$ & $0.11 \pm 0.12$ &      & 1.00 & -0.85 \\ 
$U$ & $0.14 \pm 0.09$ &      &      & 1.00  \\ \hline
\end{tabular}
  \caption{Values of the $S$, $T$, and $U$ parameters along with their correlation matrix allowed by the electroweak fit including the $W$ boson mass as measured by the CDF collaboration~\cite{Lu:2022bgw}.}
  \vspace{0.3cm}
  \label{tab:stuFit}
\end{table}
Before the new CDF measurement of the $W$ mass, the values allowed for $S$, $T$, and $U$ from the PDG electroweak fit~\cite{ParticleDataGroup:2020ssz}, $S = -0.01 \pm 0.10, T = 0.03 \pm 0.12$ and $U = 0.02 \pm 0.11$ were in good agreement with the SM prediction $S = T = U = 0$. However, the result of the CDF collaboration~\cite{CDF:2022hxs}
\begin{equation}
    M_W = \left(80.433 \pm 0.0064_\text{stat} \pm 0.0069_\text{syst}\right)~\text{GeV}
\end{equation}
challenges this electroweak fit, as it deviates by $7 \sigma$ from the SM prediction,
\begin{equation}
    \left(M_{W}\right)_\text{SM} = \left(80.379 \pm 0.012\right)~\text{GeV} ,
\end{equation}
requiring new values for $S$, $T$, and $U$. Since the electroweak parameters in the SM are closely related to each other, any modification of the $W$ mass can easily alter the other precisely measured parameters. Therefore, a careful analysis that encompasses all data is required to include the CDF measurement consistently. A dedicated fit has been performed in Ref.~\cite{Lu:2022bgw}. Here, we quote their results for the best-fit values of $S$, $T$, $U$ and their correlations in Table~\ref{tab:stuFit}. Compared to the previous electroweak fit, there is a preference for larger values for $S$, $T$, and $U$, especially in the case of the $U$ parameter.
Using these values, we determine the allowed region in the $S$, $T$, $U$ space, and, using Eqs.~\eqref{eq:stu} to~\eqref{eq:stu3}, we map it into the parameter space of our model. 
The result is shown in Fig.~\ref{stuCDFplot}, where we plot the $1$-$\protect\sigma$ and $2$-$\protect\sigma$ favored regions in the $m_{\Phi_1} - m_{\protect\eta^+}$ versus $m_{\protect\eta^+}$ plane. In the panel on the left (right), the mass of the neutral scalar $\Phi_2$ is $m_{\Phi_2} = 1$~TeV ($m_{\Phi_2} = 2$~TeV), while the mixing angle is fixed at $\protect\theta_\Phi = 0.2$ in both plots.

We note that the CDF measurement can accommodate scalar masses in the TeV scale, provided that the mass splitting among the charged and neutral scalars is not larger than a few hundred GeV. In this scenario, in addition to explaining the anomaly and avoiding the current experimental constraints, the model becomes highly predictive. To illustrate this aspect, in the plots we show two benchmark points (BP) that provide clear signatures in future flavor-violation experiments, as detailed in the next section.

\subsection{Charged lepton flavor violating observables}
In what follows, we will derive constraints arising from the charged lepton flavor violating (cLFV) decays $\mu \rightarrow e\gamma $, $\mu \to e e e $ and from the conversion $\mu-e$ in atomic nuclei, which arise at one-loop level from the virtual exchange of the electrically charged scalars $\eta ^\pm$, originating from the $SU(2)_L$ inert doublet $\eta$ and the RH Majorana neutrinos $N_{R_i}$ ($i=1, 2$).

The branching ratio for the $l _\alpha \to l _\beta \gamma$ decays is given by~\cite{Ma:2001mr, Toma:2013zsa, Vicente:2014wga, Lindner:2016bgg}
\begin{equation} \label{BRclfv}
    \text{BR}\left(l_\alpha \to l_\beta \gamma\right) = \frac{3 \left(4 \pi\right)^3 \alpha_\text{EM}}{4\, G_F^2} \left\vert A _D \right\vert^2\, \text{BR}\left(l_\alpha \to l_\beta \nu_\alpha \overline{\nu_\beta}\right) ,
\end{equation}
where the form factor $A_D$, which arises from dipole-photon penguin diagrams, takes the form
\begin{equation}
    A_D = \sum_{i=1}^2 \frac{z_{i\beta}^*\, z_{i\alpha}}{2 (4 \pi)^2} \frac{1}{m_{\eta^+}^2} F_2\left(\xi_i\right) .
\end{equation}
Here $z_{is} = \sum_{k=1}^3 y_\eta^{ks} \left(V_{lL}^\dagger\right)_{ik}$, where $V_{lL}$ is the rotation matrix that diagonalizes $M_lM^{\dagger}_l$ the charged lepton mass matrix,\footnote{Note that $z _{is}$ reduces to $y _\eta ^{is}$ in the case that the charged lepton mass matrix is diagonal.}  and $\xi _i = m _{N _{R _i}} ^2 / m _{\eta ^+} ^2$, with $m _{\eta ^+}$ being the mass of the charged scalar component of the $SU(2)_L$ inert doublet, while $m _{N _{R _i}}$ correspond to the masses of the RH Majorana neutrinos.
On the other hand, the branching ratio for 3-body decays $\ell_\alpha \to 3 \, \ell_\beta$ is given by~\cite{Toma:2013zsa, Vicente:2014wga}
\begin{align}
    \text{BR}\left(\ell_\alpha \to \ell_\beta \overline{\ell_\beta} \ell_\beta\right) &= \frac{3 (4\pi)^2 \alpha_\mathrm{EM}^2}{8\, G_F^2} \left[|A_{ND}|^2 + |A_D|^2 \left(\frac{16}{3} \log\left(\frac{m_\alpha}{m_\beta}\right) - \frac{22}{3}\right)\right. + \frac16 |B|^2 \nonumber\\
    & \quad \left. + \left(-2 A_{ND}\, A_D^* + \frac13 A_{ND} B^* - \frac23 A_D B^* + \mathrm{H.c.}\right)\right] \text{BR}\left(\ell_\alpha \to \ell_\beta \nu_\alpha \overline{\nu_\beta}\right). \label{eq:l3lBR}
\end{align}
The form factor $A_{ND}$ is generated from the non-dipole photon penguin diagrams and takes the form
\begin{equation} \label{eq:AND}
    A_{ND}=\sum_{i=1}^3\frac{z_{i\beta}^*\, z_{i\alpha}}{6 (4 \pi)^2} \frac{1}{m_{\eta^+}^2} G_2\left(\xi_i\right).
\end{equation}
In addition to that, the contribution of box diagrams generates the form factor $B$, which is given by
\begin{equation} \label{eq:B}
    e^2 B = \frac{1}{(4 \pi)^2 m_{\eta^+}^2} \sum_{i,\, j = 1}^3 \left[\frac12 D_1(\xi_i,\, \xi_j)\, z_{j \beta}^*\, z_{j \beta}\, z_{i \beta}^*\, z_{i \alpha} + \sqrt{\xi_i\, \xi_j} D_2(\xi_i,\, \xi_j)\, z_{j \beta}^*\, z_{j \beta}^*\, z_{i \beta}\, z_{i \alpha}\right],
\end{equation}
where the different loop functions take the form
\begin{align}
    F_2(x) &= \frac{1 - 6 x + 3 x^2 + 2 x^3 - 6 x^2 \log x}{6 (1-x)^4}\,,\\
    G_2(x) &= \frac{2 - 9 x + 18 x^2 - 11 x^3 + 6 x^3 \log x}{6 (1-x)^4}\,, \\
    D_1(x,y) &= -\frac{1}{(1-x) (1-y)} - \frac{x^2 \log x}{(1-x)^2 (x-y)} - \frac{y^2 \log y}{(1-y)^2 (y-x)}\,, \\
    D_2(x,y) &= -\frac{1}{(1-x) (1-y)} - \frac{x \log x}{(1-x)^2 (x-y)} - \frac{y \log y}{(1-y)^2 (y-x)}\,.
\end{align}

Note that the large factor in front of $\vert A _D \vert ^2$ in Eq.~\eqref{eq:l3lBR} makes the contribution of the form factor $A_D$ to be typically more important than the contribution of $A_{ND}$ in the $\mu \to eee$ decay. Also note that the form factor $B$ is proportional to the fourth power of the Yukawa couplings, while $A _D$ and $A _{ND}$ are proportional to the second power. Therefore, when the Yukawa couplings are sufficiently small, the contribution $A _D$ tends to dominate the amplitude. In this case, the processes $\mu \to e e e$ and $\mu \to e \gamma$ are correlated in a simple way, \textit{i.e.}, the rate $\mu \to e e e$ becomes proportional to $\mu \to e \gamma$ with a proportionality factor much smaller than one. This scenario of dipole dominance is the typical case studied in the literature, in which the decay of $\mu \to e e e$ is suppressed with respect to $\mu \to e \gamma$, and the strongest constraints come from the process $\mu \to e \gamma$. However, in the regime of large Yukawa couplings, the box contributions become sizable and cannot be neglected, especially in the limits $m _{\eta ^+} \gg m _{N _R}$ or $m _{\eta ^+} \ll m _{N _R}$, in which $B$ is enhanced compared to $A _D$, due to the particular behavior of the loop functions $D _1$, $D_2$, and $F _2$. In this case, the decay $\mu \to eee$ yields competitive bounds.

\begin{figure}[t!] 
    \includegraphics[width=0.48\linewidth]{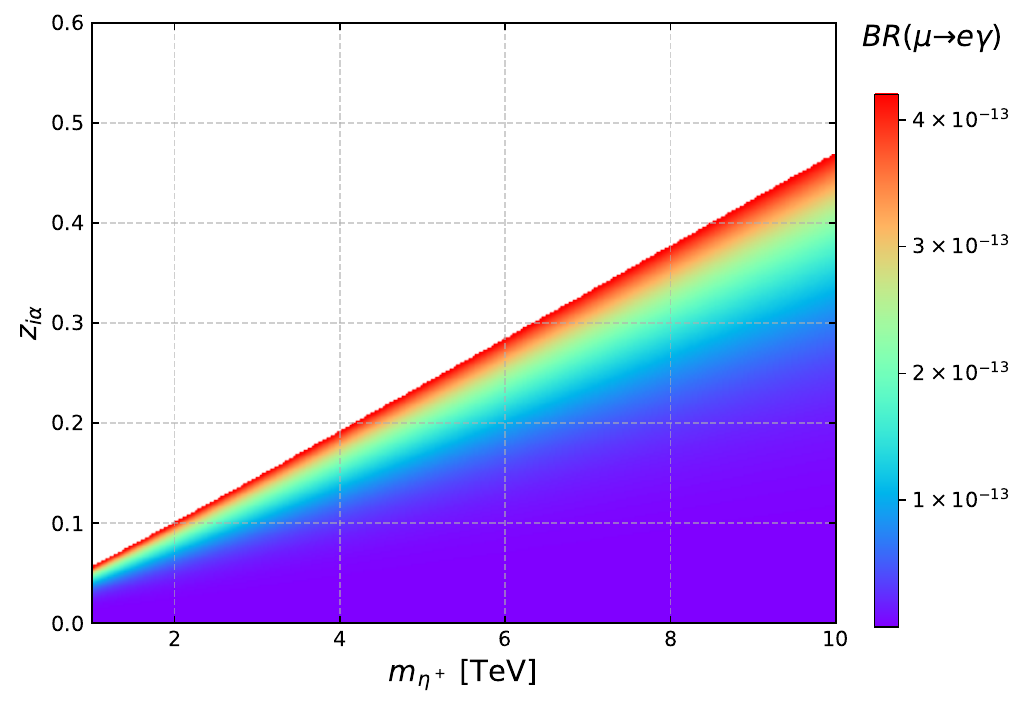} \quad \includegraphics[width=0.48\linewidth]{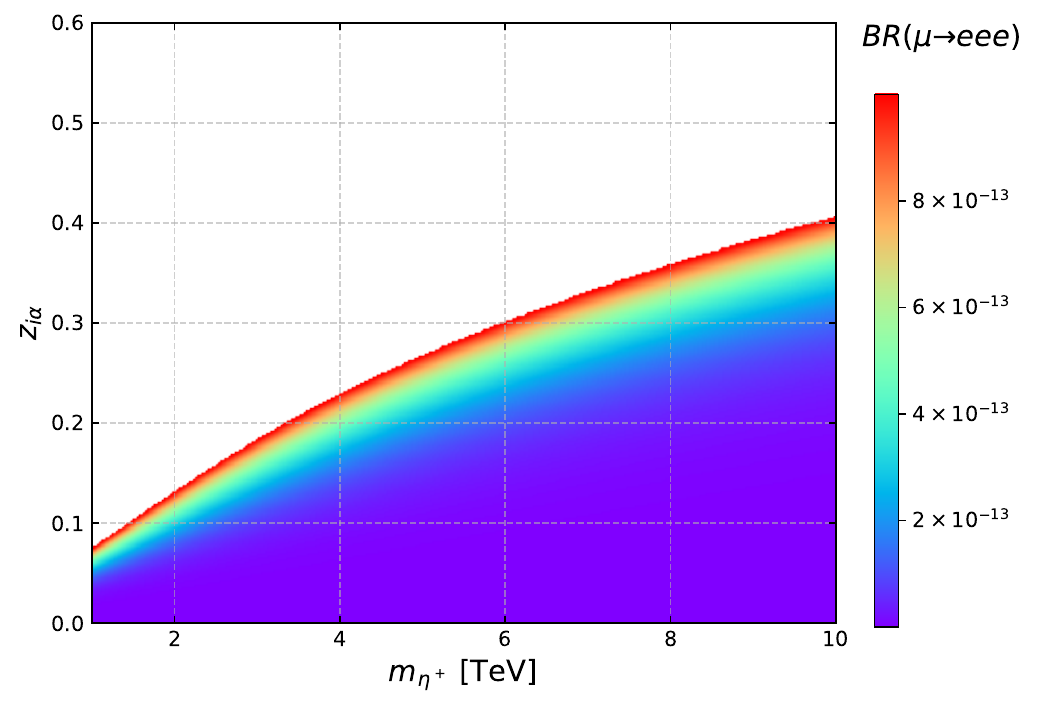}  
    \caption{Parameter space in the $m _{\eta ^+} - z _{i \alpha}$ plane consistent with the charged lepton flavor violation limits. The colored regions are allowed by the current constraints.}
    \label{LFVplot2}
\end{figure}
This is illustrated in Fig.~\ref{LFVplot2}, where the allowed parameter space in the $m _{\eta ^+}-z_{i \alpha}$ plane is shown after imposing the bounds from MEG~\cite{MEG:2016leq} on $\mu \to e \gamma$ (left panel) and SINDRUM~\cite{SINDRUM:1987nra} on $\mu \to e e e$ (right panel). We consider the Yukawa couplings of $\mathcal{O}(0.1)$, which is high enough for the box contributions to be relevant. For simplicity, we assumed that all Yukawa couplings are equal and degenerate Majorana neutrinos, with a fixed mass $m_{N_R} = 500$~GeV. One can notice that the excluded parameter space for both processes is very similar, with $\mu \to eee$ becoming slightly more constraining for larger $m_{\eta^+}$ values (due to the enhancement of the box contributions in the regime $m _{\eta ^+} \gg m _{N _R}$). One can also see that for Yukawa couplings of order $\mathcal{O}(0.1)$, the model complies with the cFLV bounds provided that $\eta ^+$ and $N_{R_i}$ have masses in the TeV ballpark. On the other hand, if the couplings and masses lie around this range, cLFV rates can be sufficiently high to be within the reach of future experiments, such as MEG II and Mu3e, as we discuss next.

Before that, let us turn to the conversion $\mu-e$ in the nuclei, whose ratio $\mu^{-}-e^{-}$ is defined as~\cite{Lindner:2016bgg}
\begin{equation}  \label{eq:Conversion-Rate}
    \text{CR}\left(\mu - e\right) = \frac{\Gamma\left(\mu^- + \text{Nucleus} \left(A,\, Z\right) \to e^- + \text{Nucleus}\left(A,\, Z\right)\right)}{\Gamma\left(\mu^- + \text{Nucleus}\left(A,\, Z\right) \to \nu_\mu + \text{Nucleus}\left(A,\,Z-1\right)\right)}\,.
\end{equation}
For the radiative neutrino mass model considered in this work, the conversion rate, normalized to the charged lepton capture rate, takes the form~\cite{Vicente:2014wga}: 
\begin{align}
    \mathrm{CR}\left(\ell_\alpha\, N \to \ell_\beta \, N\right) &= \frac{p_\beta E_\beta m_\alpha^3 G_F^2 \alpha_\text{EM}^3 Z_\text{eff}^4 F_p^2}{8 \pi^2\, Z\, \Gamma_\text{capt}}\left[\left|(Z + N) \left(g_{LV}^{(0)} + g_{LS}^{(0)}\right) + (Z - N) \left(g_{LV}^{(1)} + g_{LS}^{(1)}\right)\right|^2 \right. \nonumber \\
    &\qquad \left. + \left|(Z + N) \left( g_{RV}^{(0)} + g_{RS}^{(0)}\right) + (Z - N) \left(g_{RV}^{(1)} + g_{RS}^{(1)}\right)\right|^2\right],
\end{align}
where the expressions for the quantities $Z_\text{eff}$, $F_p$, $\Gamma_\text{capt}$, and $g^{(i)}_{L/R \, S/V}$ are given in Refs.~\cite{Arganda:2007jw, Vicente:2014wga}.

Compared to $\mu \to e e e$ and $\mu \to e \gamma$, the conversion $\mu-e$ in nuclei currently provides very weak constraints. A similar exclusion region to that of Fig.~\ref{LFVplot2} for the $\mu-e$ conversion would exclude almost no parameter space and, therefore, is not shown. However, given that future experiments, such as Mu2e and COMET~\cite{Bernstein:2013hba}, are expected to measure or at least constrain lepton flavor conversion in nuclei with much better precision than radiative cLFV decays, we included flavor conversion processes and the other cLFV channels when analyzing their detection prospects in future cLFV experiments.

\begin{figure}[t!]
    \centering
    \includegraphics[width=0.62\linewidth]{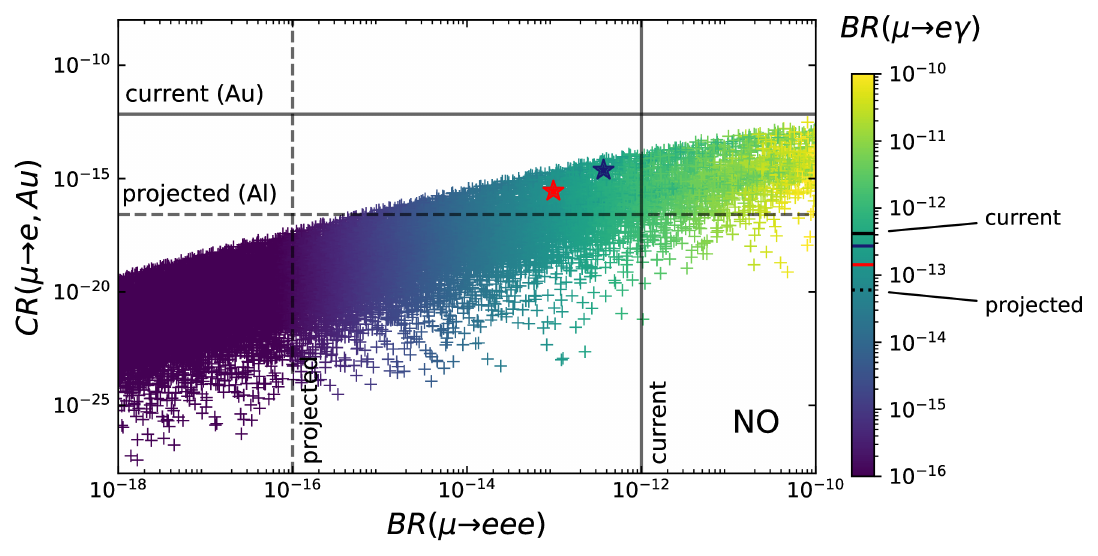}\quad 
    \includegraphics[width=0.62\linewidth]{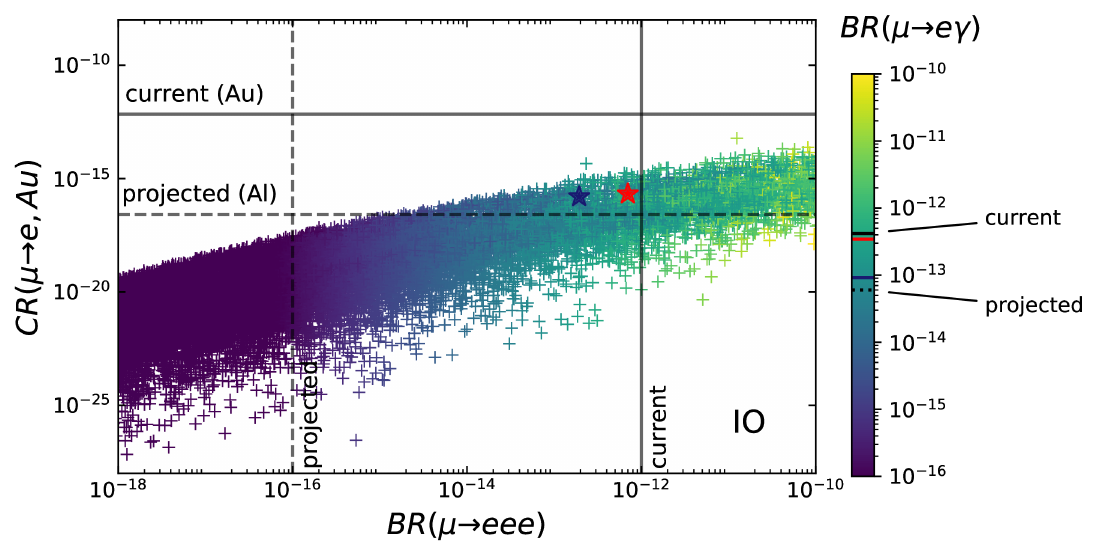}
    \caption{Correlation between the cLFV processes $\mu \to eee$ and $\mu \to e$ conversion in gold nuclei for normal (upper panel) and inverted (lower panel) neutrino mass ordering. The points are colored according to the size of the branching ratio of the $\mu \to e \gamma$ process, as shown in the color bars. The current upper bounds are indicated by the black full lines, while the future sensitivities, by the black dashed lines. The blue and red star points correspond to the benchmarks BP1 and BP2, respectively (see Table~\ref{tab:bencmarks}). In the bars, the current and projected bounds for $\mu \to e \gamma$ and the benchmark points are indicated by strips, which employ the same color convention as before.}
    \label{LFVplot3}
\end{figure}
\begin{table}[!t]
  \centering
  \begin{tabular}{|c|c|}
  \hline
\textbf{Parameters}     & \textbf{Scanned ranges} \\ \hline
$\theta _R$             & $[0, 2 \pi]$ \\
$\lambda _{14}$         & $[0.01, 1]$ \\  
$m _{N _R}$, $m_{\eta ^+}$, $m _{\Phi _{1,2}}$, $m _{\Xi _{1,2,3,4}}$           & $[500, 10000]$~GeV \\ 
\hline
\end{tabular}
  \caption{Scanned parameter ranges.}
  \vspace{0.3cm}
  \label{tab:scanRange}
\end{table}
We perform a scan over the parameter space of the model and calculate the corresponding cLFV rates. The result is shown in Fig.~\ref{LFVplot3}, for both neutrino mass orderings, NO (top panel) and IO (bottom panel). The input parameters are randomly scanned according to the ranges shown in Table~\ref{tab:scanRange}. The scalar mixing angles were fixed as $\theta _\Phi = 0.2$, $\theta _\Xi = 0.3$ and ${\theta _\Xi} ^\prime = 0.1$. For simplicity, we assumed that RH neutrinos are degenerate with mass ($m _{N _R}$). 
We chose $\lambda _{14}$ in the range indicated in Table~\ref{tab:scanRange}, as we realized that $\lambda _{14}$ in this range allows Yukawa couplings of $\mathcal{O} (0.1)$; larger $\lambda _{14}$ values would decrease the Yukawas and suppress the cLFV rates.
Following the recipe of Section~\ref{Sec:Neutrino} for the adapted Casas-Ibarra parameterization, shown in Eq.~\eqref{eq:adaptedCasasIbarra}, these parameters enter as inputs for the Yukawa couplings, which depend on $\lambda _{14}$ and on the scalar masses and mixing angles through the $\Lambda$ function (cf. Eq.~\eqref{eq:lambda}), and also on $\theta _R$, the arbitrary angle of the orthogonal matrix $R$ (which we assume to be real). In this manner, the neutrino masses and mixing are guaranteed to be reproduced for all the calculated points.

We observe from Fig.~\ref{LFVplot3} that, indeed, the current limits from the $\mu-e$ conversion in the nuclei are very weak and practically do not give any constraints on the model. However, it is apparent that the high precision expected for future $\mu-e$ conversion experiments will allow a large portion of the parameter space of the model to be probed. In the plots, the $\text{CR}(\mu \to e)$ values were calculated considering the gold (Au) nucleus to be compared with the current bound from the SINDRUM experiment, which used this nucleus as the target. However, note that the projected sensitivity refers to the aluminum (Al) nucleus, which is used as the target in the COMET and Mu2e experiments. However, we checked that the difference in the predicted rates for the Au and Al nuclei is only marginal along the parameter space of interest. Therefore, a comparison of the projected bounds involving these two nuclei is possible.
\begin{table}[!t]
  \centering
  \begin{tabular}{|c | c c | c c|}
  \hline
\textbf{Parameters}   & \multicolumn{2}{c|}{\textbf{BP1}} & \multicolumn{2}{c|}{\textbf{BP2}} \\ \hline
$\theta _\Phi$        & \multicolumn{2}{c|}{$0.2$}        & \multicolumn{2}{c|}{$0.2$} \\
$\theta _\Xi$         & \multicolumn{2}{c|}{$0.3$}        & \multicolumn{2}{c|}{$0.3$} \\
$\theta _\Xi ^\prime$ & \multicolumn{2}{c|}{$0.1$}        & \multicolumn{2}{c|}{$0.1$} \\
$m _{\eta ^+}$ [GeV]  & \multicolumn{2}{c|}{$1500$}       & \multicolumn{2}{c|}{$1700$} \\ 
$m _{\Phi _1}$ [GeV]  & \multicolumn{2}{c|}{$1600$}       & \multicolumn{2}{c|}{$1765$} \\ 
$m _{\Phi _2}$ [GeV]  & \multicolumn{2}{c|}{$1000$}       & \multicolumn{2}{c|}{$2000$} \\ \hline
 & \textbf{\footnotesize{NO}} & \textbf{\footnotesize{IO}} & \textbf{\footnotesize{NO}} & \textbf{\footnotesize{IO}} \\ \hline
$m _{N _R}$ [GeV]       & $ 8954.5$   & $ 4246.9$     & $ 5040.0$  & $ 3450.7$ \\ 
$m _{\Xi _1}$ [GeV]     & $ 8130.4$   & $ 2925.0$     & $ 8244.0$  & $ 3282.9$ \\ 
$m _{\Xi _2}$ [GeV]     & $ 1452.5$   & $ 4748.5$     & $ 2431.6$  & $ 1815.4$ \\ 
$m _{\Xi _3}$ [GeV]     & $ 8932.4$   & $ 2763.1$     & $ 6392.1$  & $ 3637.6$ \\ 
$m _{\Xi _4}$ [GeV]     & $ 7127.2$   & $ 9336.4$     & $ 1296.0$  & $ 1458.4$ \\ 
$\lambda _{14}$         & $ 0.729 $   & $ 0.726 $     & $ 0.363 $  & $ 0.504 $ \\ 
$y _\eta ^{e 1}$        & $ 0.124 $   & $ 0.346 $     & $-0.009 $  & $ 0.639 $ \\ 
$y _\eta ^{e 2}$        & $-0.253 $   & $ 0.389 $     & $ 0.154 $  & $ 0.152 $ \\ 
$y _\eta ^{\mu 1}$      & $ 0.746 $   & $ 0.220 $     & $-0.313 $  & $ 0.031 $ \\ 
$y _\eta ^{\mu 2}$      & $-0.307 $   & $-0.272 $     & $ 0.312 $  & $-0.440 $ \\ 
$y _\eta ^{\tau 1}$     & $ 0.705 $   & $-0.335 $     & $-0.400 $  & $-0.183 $ \\ 
$y _\eta ^{\tau 2}$     & $ 0.207 $   & $ 0.225 $     & $ 0.043 $  & $ 0.475 $ \\ 
\hline \hline
\textbf{Observables}    & \multicolumn{2}{c|}{ } & \multicolumn{2}{c|}{ } \\ \hline
$\text{BR}(\mu \to e \gamma)$ & $ 2.730 \times 10 ^{-13} $     & $ 9.170 \times 10 ^{-14} $  & $ 1.428 \times 10 ^{-13} $ & $ 3.452 \times 10 ^{-13} $ \\
$\text{BR}(\mu \to e e e)$    & $ 3.686 \times 10 ^{-13} $     & $ 1.933 \times 10 ^{-13} $  & $ 9.799 \times 10 ^{-14} $ & $ 6.997 \times 10 ^{-13} $ \\
$\text{BR}(\mu - e, Au)$      & $ 2.392 \times 10 ^{-15} $     & $ 1.599 \times 10 ^{-16} $  & $ 2.816 \times 10 ^{-16} $ & $ 2.122 \times 10 ^{-16} $ \\
$m _{ee}$ [meV]              & $ 3.67 $     & $ 48.36 $  & $ 3.67 $ & $ 48.36 $ \\ \hline
\end{tabular}
  \caption{Input parameters, cLFV observables and $0\nu 2\beta$ effective mass for the selected benchmark points BP1 and BP2. They satisfy all the current constraints and accommodate the CDF anomaly while offering detectable signals in future cLFV and $0\nu 2\beta$ decay experiments.}
  \vspace{0.3cm}
  \label{tab:bencmarks}
\end{table}

We also highlight the important role that future $\mu \to e e e$ searches play in our model. One can see that it will be able to cover a very significant part of its parameter space. This is a consequence of the high precision expected for the Mu3e experiment, combined with the high rates predicted for this process due to the box-diagram contributions. On the other hand, despite the sizable rates predicted for the $\mu \to e \gamma$ process, the expected experimental accuracy of MEG II does not allow one to probe as much parameter space as the other cLFV searches. 

In Fig.~\ref{LFVplot3} we also show two benchmark points, BP1 (blue star) and BP2 (red star). The values of the various parameters are detailed in Table~\ref{tab:bencmarks}. These points correspond to the same BP1 and BP2 shown in Fig.~\ref{stuCDFplot}, which explain the $W$ mass anomaly. However, in that plot, as the neutrino ordering does not affect the $S$, $T$, and $U$ parameters, the points in NO and IO become degenerate, and this is why they appear as single points in that figure. Note that both BP1 and BP2 provide measurable rates for the three cLFV processes considered above, in both neutrino mass orderings. In addition to that, in the specific case of IO, a $0\nu 2\beta$ decay signal would also be possible. These benchmark points are concrete examples of how this three-loop model allows one to generate active neutrino masses without fine-tuning the Yukawa couplings (or any other parameter), explain the $W$ mass anomaly, and offer new signatures in future experiments. Further cosmological features of the model, such as baryogenesis and dark matter, are discussed in the next section.

\subsection{Muon $g-2$}

\begin{figure}[t!]
\centering
\begin{subfigure}{.49\textwidth}
  \centering
  \includegraphics[width=\textwidth]{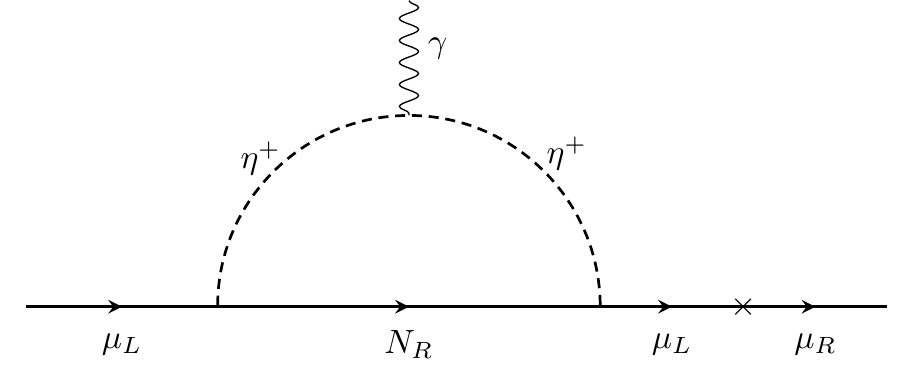} 
  \caption{}
  \label{fig:g-2_RHv}
\end{subfigure}
\begin{subfigure}{.43\textwidth}
  \centering
  \includegraphics[width=\textwidth]{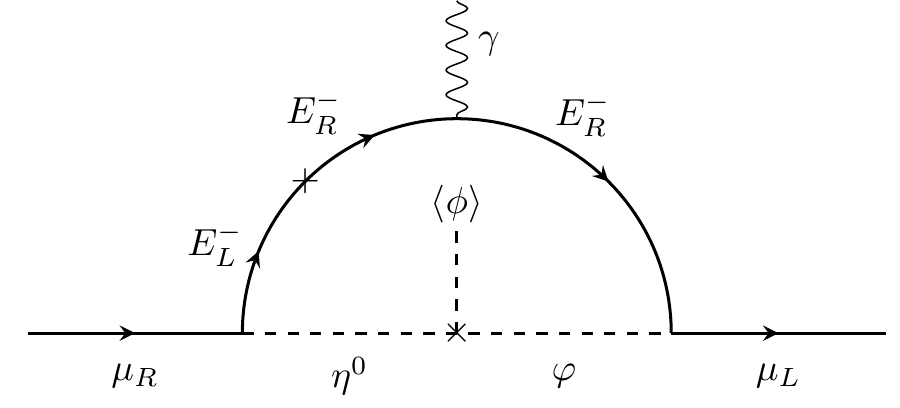} 
  \caption{}
  \label{fig:g-2_vectorlike}
\end{subfigure}
\caption{1-loop contributions to the muon anomalous magnetic moment. The diagram (a) leads to a negative value of $\Delta a _\mu < 0$. With the inclusion of a vector-like lepton $E$ in diagram (b), a positive value of $\Delta a _\mu > 0$ is predicted.}
\label{fig:g-2}
\end{figure}

There is a persisting positive deviation $\Delta a_\mu>0$ of the experimental value of the anomalous magnetic moment of muon $a_\mu = (g-2)/2$~\cite{Muong-2:2021ojo, Muong-2:2021ovs, Muong-2:2021vma} from its SM value~\cite{Aoyama:2020ynm}, which can be interpreted as a contribution of physics beyond the SM.
The model we proposed in this work predicts a negative value of $\Delta a_\mu<0$ in contrast to the above-mentioned positive-valued experimental result.
In our model, $\Delta a_\mu$ is generated according to Fig.~\ref{fig:g-2_RHv} at one-loop level by the electrically charged scalar $\eta^{+}$ and a right-handed Majorana neutrino $N_{R}$ running in the internal lines of the loop, where chirality flipping occurs only in the external fermionic line. This contribution to $\Delta a_\mu$ is known to be negative~\cite{Lindner:2016bgg}. 
To resolve this issue, our model should be properly modified. This topic 
is beyond the scope of the present paper and should be addressed elsewhere.  Here, only a brief comment is in order. Our model can be minimally extended by adding exotic $SU(2)_L$-singlet vector-like leptons $E_{L,R}$ with negative $\mathbb{Z}_2$-parity, weak hypercharge $Y=-1$ and the $U'(1)$ charge $Q'=-6$.
With this assignment, we have new Lagrangian terms
\begin{eqnarray} \label{eq:E-LR}    
\mathcal{L} \supset y_{\eta E}\,\overline{l}_{2L} \eta E_R + y_{\varphi E}\,\overline{E}_L \varphi^*\mu_R + M_E \overline{E}_LE_R+h.c.
\end{eqnarray}
obeying all the symmetries of our model. The mass $M_E$ of the exotic leptons should be large enough to pass the existing experimental limits~\cite{Morais:2021ead}.
These new terms together with the trilinear coupling $A(\eta^{\dagger }\phi )\varphi$
in the scalar potential of Eq.~\eqref{eq:sacalar-potential} allow us to construct an extra one-loop contribution to $\Delta a_\mu$ shown in Fig.~\ref{fig:g-2_vectorlike}. In this case, the photon is attached to the charged internal $E^-$ line and the helicity-flip is due to large $M_E$. This contribution is positive and can easily be made larger than the negative contribution induced 
by the $\eta^+ - N_R$ loop we mentioned earlier.
In this minimal way, we can successfully accommodate the experimental magnitude and sign of $\Delta a_\mu$ without having affected all the results derived in the present paper.

\section{Cosmological implications} \label{Sec:cosmo}
In this section, a couple of phenomenological implications for early universe cosmology are discussed: the domain-wall issue, possible production mechanisms for the DM, and the baryon asymmetry of the universe.

\subsection{\boldmath The $U'(1)$ Goldstone and Domain Walls}
In our model, there is a potential cosmological issue related to the presence of the global $U'(1)$ symmetry spontaneously broken at the scale $v_{\sigma} \geq 1$~TeV, which is significantly larger than the electroweak scale $v\sim 200$~GeV. 
First, this entails the appearance of a massless electroweak singlet Goldstone boson, which can acquire its small mass due to the Planck-scale effects of quantum gravity~\cite{Akhmedov:1992hi, Rothstein:1992rh}. This particle, similar to a singlet majoron, is phenomenologically harmless. On the other hand, it could serve as a DM candidate if its mass was in the keV range~\cite{Akhmedov:1992hi, Lattanzi:2007ux}, which seems to be too large for Planck-scale effects and would require soft breaking $U'(1)$ symmetry. We do not consider this option and leave $U'(1)$ Goldstone without further discussion.

The issue arises from the side of the domain-wall formation.
We note that non-perturbative instanton effects~\cite{tHooft:1976rip}, originating from the mixed $[SU(2)_L]^2\times U'(1)$ anomaly, explicitly break this symmetry $U'(1)$ down to a discrete $\mathbb{Z}_N$. According to Ref.~\cite{Lazarides:2018aev} and our $U'(1)$ assignments in Table~\ref{Themodel}, we have 
\begin{equation} \label{eq:instanton}
    N = \left|\sum_R N(R)\times Q'(R)\times T(R)\right| = \left|\left(3\times 3\right)\times \frac{1}{3}\times 1 - 3\times 3 \times 1 \right| = 6\,,
\end{equation}
where $N(R)$ is the number of a fermion representation $R$, the $U'(1)$ charge of the fermion is denoted by $Q'(R)$, and $T(R)$ is the Dynkin index of the group $SU(2)_L$, which is $T(1)=0$ and $T(2)=1$ for singlets and doublets, respectively. In Eq.~\eqref{eq:instanton} we explicitly show two contributions that arise in our model. The first term is the quark-doublet contribution with $N(q_L)= (\text{number of generations})\times (\text{number of colors})=3\times 3$, $Q'(q_L)=1/3$, while the second term is the contribution of lepton doublets with $N(\ell_L)= (\text{number of generations})=3$ and $Q'(\ell_L)=-3$.
Thus, the instantons break $U'(1)$ down to its subgroup $\mathbb{Z}_6$. 
However, spontaneous breaking is induced by a unique term $\lambda_{15}\, \rho\, \zeta\, \sigma^{2}$ in the scalar potential in Eq.~\eqref{eq:sacalar-potential} leading to $\Delta Q' = 1$ breaking. Then $U'(1)$ is spontaneously broken down to the trivial $\mathbb{Z}_1$.  Therefore, the exact $\mathbb{Z}_6$ is broken spontaneously, leading to the formation of the domain walls. There are several mechanisms proposed in the literature to solve this problem from the cosmological and particle physics sides. In the first case, one appeals to a suitable inflation scenario, in which the dangerous domain walls are beyond the horizon. The most known post-inflation particle physics solution resorts to the Lazarides-Shafi mechanism~\cite{Lazarides:1982tw} which assigns the spontaneously broken discrete symmetry -- in our case, this is $\mathbb{Z}_6$ -- to be a subgroup of some gauge group. This allows one to gauge away the physical degeneracy between the different vacua.  
The above two mechanisms can be implemented in our model without modifying its field content and symmetries.
Invoking the Lazarides-Shafi mechanism would imply that the global $U'(1)$ is a subgroup of some gauge group $G$ so that $\mathbb{Z}_6\in U'(1)$ is the center of $G$. The group $G$ is supposed to be spontaneously broken at a sufficiently high scale, down to a group containing the global $U'(1)$ subgroup. We emphasize that the choice of a particular mechanism from these two does not affect the results of the present study.  

\subsection{Baryon asymmetry of the Universe}
The baryon asymmetry of the Universe may be generated by a combination of RH neutrino out-of-equilibrium decays and the sphaleron process, similar to the canonical (TeV scale) seesaw mechanism~\cite{Hambye:2001eu}. 
In our three-loop neutrino mass model, the size of the Yukawa neutrino coupling is as large as $\mathcal{O}(0.1)$ as shown in Table~\ref{tab:bencmarks}, leading to a strong washout effect through scattering processes such as $\ell_i\, \eta^+ \leftrightarrow \bar{\ell}_i\, \eta^-$ and $\ell_i\ell_i\leftrightarrow \eta^-\eta^-$. 
The size of the Yukawa neutrino coupling should be smaller than $\mathcal{O}(10^{-4})-\mathcal{O}(10^{-3})$ for successful leptogenesis on the TeV scale~\cite{Kashiwase:2012xd, Borah:2018rca}. 
Therefore, it is difficult to produce baryon asymmetry through leptogenesis in this scenario.\footnote{A recent study considers a scalar mass lighter than the sphaleron process freeze-out temperature, which could be a possible way to avoid the strong wash-out effect~\cite{Seto:2022tow}.}
Inert scalar annihilations may also be used to generate lepton asymmetry instead of RH neutrino decays. 
This framework is named WIMPy leptogenesis~\cite{Ahriche:2017iar}. 
Even in this case, the washout effect is strong enough to erase the produced lepton asymmetry. 

A possible idea for generating the baryon asymmetry in our model may be electroweak baryogenesis. 
Electroweak baryogenesis in a different three-loop neutrino mass model has been studied in Refs.~\cite{Aoki:2008av, Aoki:2009vf}. 
If a CP-violating phase exists in the scalar potential and a strong electroweak first-order phase transition occurs, the baryon asymmetry may be generated via electroweak baryogenesis in our model, although this is not the scope of this work. 

\subsection{Dark Matter}
First, we note that the $\mathbb{Z}_2$ symmetry protects the lightest odd state of the dark sector, making it a viable candidate for DM. In the present scenario, one could have a scalar ($\eta^0$, $\varphi$, $\rho$ or $\zeta$) or fermionic ($N_{R_1}$) DM, depending on the mass hierarchy.
For DM masses in the GeV to TeV ballpark and couplings to the SM at the electroweak scale, DM could have been generated in the early universe via the WIMP mechanism.
As it has been previously shown that this model offers a quite broad viable parameter space, instead of performing an intensive scan to find the precise regions compatible with the observed DM abundance, here we prefer to give some general comments on WIMP DM phenomenology.

The case of scalar DM corresponds to the scenario in which the candidate for DM is the lightest of $\eta^0$, $\varphi$, $\rho$, and $\zeta$. 
In the early universe, the DM candidate can be produced out of the SM thermal bath by the WIMP mechanism via a number of 2-to-2 processes. A pair of DM particles can annihilate into two SM states via the $s$ channel exchange of a Higgs boson or to a couple of Higgs bosons due to the contact interaction or the $t$- and $u$-channel exchange of a DM. Furthermore, it can also annihilate into a couple of active neutrinos by the mediation of a RH neutrino in the $t$- and $u$-channels.
If the dominant component of DM is $\eta^0$, the properties of DM are similar to those of the inert scalar DM~\cite{LopezHonorez:2006gr}. 
The main annihilation channels are into the gauge bosons and Higgs bosons via the gauge interactions. 
On the other hand, if DM is dominated by the other components, the main annihilation channels are into the Higgs bosons via the scalar couplings. This scenario corresponds to the Higgs portal DM~\cite{McDonald:1993ex, Burgess:2000yq}. 

Alternatively, for fermionic DM, the candidate is the lightest state $N_{R_1}$. 
In the early Universe, it can be annihilated into a couple of charged leptons or active neutrinos via the $\eta$ exchange ($t$- and $u$-channels). 
The DM properties for this scenario are similar to those of the original scotogenic model~\cite{Ma:2006km}, and a detailed study has been carried out in Refs.~\cite{Kubo:2006yx, Suematsu:2009ww, Schmidt:2012yg, Cacciapaglia:2020psm, Rosenlyst:2021tdr}. 

Before closing this section, it is interesting to note that in the present scenario, there are alternatives to the WIMP mechanism. For example, DM could also have been produced non-thermally in the early Universe via the FIMP mechanism~\cite{McDonald:2001vt, Choi:2005vq, Kusenko:2006rh, Petraki:2007gq, Hall:2009bx, Bernal:2017kxu}. In that case, much smaller portal couplings $\mathcal{O}(10^{-11})$ with a wider mass range are required. Furthermore, if the portal couplings are suppressed and DM has significant self-interactions, thermalization in the dark sector could have a strong impact on the DM abundance~\cite{Bernal:2015xba, Heikinheimo:2017ofk, Bernal:2020gzm, Arcadi:2019oxh, DeRomeri:2020wng, Herms:2018ajr, Hansen:2017rxr}. Finally, we note that non-standard cosmological histories could also greatly modify the expected DM abundance, with respect to the standard scenario~\cite{Bernal:2018ins, Hardy:2018bph, Bernal:2018kcw, Allahverdi:2020bys}.
 
\section{Conclusions} \label{Sec:conclusions}
We have constructed an extended Inert Doublet model (IDM) with a spontaneously broken global symmetry $U(1)'$ and a preserved $\mathbb{Z}_2$ symmetry, which allows the implementation of a three-loop radiative seesaw mechanism that produces the tiny active neutrino masses. In our proposed model, the scalar and fermion sectors are enlarged by the inclusion of four electrically neutral gauge singlet scalars and two right-handed Majorana neutrinos, respectively. In this setup, a fermionic or scalar DM candidate can easily be accommodated, whose stability is ensured by the preserved $\mathbb{Z}_2$ symmetry. Moreover, the three-loop suppression allows the new particles to have masses in the TeV scale without fine-tuning the Yukawa couplings. With the new particles in that scale, we have shown that the model is capable of successfully explaining the $W$ mass anomaly, provided that the mass splitting between the charged and neutral inert scalars is not too large. Furthermore, the model leads to interesting phenomenology while satisfying all the current constraints imposed by neutrinoless double-beta decay, charged-lepton flavor violation, and electroweak precision observables. In particular, the relatively large Yukawa couplings lead to sizable rates for the charged lepton flavor violation processes which are within the sensitivity of future facilities, especially those looking for three-body decay $\mu \to e e e$ and $\mu - e$ conversion in atomic nuclei, rendering the model testable in the next generation of flavor violation experiments.

\section*{Acknowledgments}
NB received funding from the Spanish FEDER/MCIU-AEI under grant
FPA2017-84543-P. AECH is supported by ANID-Chile FONDECYT 1210378, 1190845, ANID PIA/APOYO AFB220004, and ANID Programa Milenio code ICN2019$\_$044. TBM acknowledges CNPq (grant No. 164968/2020-2) and ANID-Chile FONDECYT (grant No. 3220454) for financial support.
This project has received funding and support from the European Union's Horizon 2020 research and innovation programme under the Marie Sk{\l}odowska-Curie grant agreement No.~860881 (H2020-MSCA-ITN-2019 HIDDeN). 
This work was supported by the JSPS Grant-in-Aid for Scientific Research KAKENHI Grant No. JP20K22349 (TT).

\appendix
\section{Tree level stability conditions of the scalar potential}
Here, we determine the stability conditions of the scalar potential. To this end, we proceed to analyze its quartic terms, since they will provide the leading contribution to the behavior of the scalar potential in a regime of very large values of the field components. The quartic terms of the model scalar potential are given by\footnote{Here, in order to simplify our analysis we have considered the case of $\kappa_i=0$, ($i=1,2,\dots 7$).}
\begin{eqnarray}
V_{4} &=&\lambda _{1}(\phi ^{\dagger }\phi )^{2}+\lambda _{2}(\sigma ^{\ast
}\sigma )^{2}+\lambda _{3}(\phi ^{\dagger }\phi )(\sigma ^{\ast }\sigma
)+\lambda _{4}(\eta ^{\dagger }\eta )^{2}+\lambda _{5}(\varphi ^{\ast
}\varphi )^{2}  \notag \\
&&+\lambda _{6}(\rho ^{\ast }\rho )^{2}+\lambda _{7}(\zeta ^{\ast }\zeta )^{2}+\lambda
_{8}(\eta ^{\dagger }\eta )(\varphi ^{\ast }\varphi )+\lambda _{9}(\eta
^{\dagger }\eta )(\rho ^{\ast }\rho )+\lambda _{10}(\eta ^{\dagger }\eta)(\zeta ^{\ast }\zeta )  \notag \\
&&+\lambda _{11}(\varphi ^{\ast }\varphi )(\rho ^{\ast }\rho )+\lambda
_{12}(\varphi ^{\ast }\varphi )(\zeta ^{\ast }\zeta )+\lambda _{13}(\rho ^{\ast }\rho
)(\zeta ^{\ast }\zeta )+\left(\lambda _{14} \varphi \rho ^{3}+\mathrm{H.c.}\right)  \notag \\
&&+\left(\lambda _{15}\rho \zeta \sigma ^{2}+\mathrm{H.c.}\right) +\lambda _{16}(\phi
^{\dagger }\phi )(\eta ^{\dagger }\eta )+\lambda _{17}(\phi ^{\dagger }\eta
)(\eta ^{\dagger }\phi )+\lambda _{18}(\phi ^{\dagger }\phi )(\varphi ^{\ast
}\varphi )  \notag \\
&&+\lambda _{19}(\phi ^{\dagger }\phi )(\rho ^{\ast }\rho )+\lambda
_{20}(\phi ^{\dagger }\phi )(\zeta ^{\ast }\zeta )+\lambda _{21}(\sigma ^{\ast }\sigma
)(\eta ^{\dagger }\eta )+\lambda _{22}(\sigma ^{\ast }\sigma )(\varphi
^{\ast }\varphi )  \notag \\
&&+\lambda _{23}(\sigma ^{\ast }\sigma )(\rho ^{\ast }\rho )+\lambda
_{24}(\sigma ^{\ast }\sigma )(\zeta ^{\ast }\zeta ).
\end{eqnarray}
We introduce the following bilinear combinations of the scalar fields
\begin{eqnarray}
a &=&\phi ^{\dagger }\phi ,\hspace{0.5cm}\hspace{0.5cm}b=\sigma ^{\ast
}\sigma ,\hspace{0.5cm}\hspace{0.5cm}c=\eta ^{\dagger }\eta ,\hspace{0.5cm}%
\hspace{0.5cm}d=\varphi ^{\ast }\varphi , \notag\\
e &=&\rho ^{\ast }\rho ,\ \hspace{0.5cm}\hspace{0.5cm}f=(\zeta ^{\ast }\zeta )\ ,\hspace{%
0.5cm}\hspace{0.5cm}r=\phi ^{\dagger }\eta +\eta ^{\dagger }\phi ,\hspace{%
0.5cm}\hspace{0.5cm}s=-i\left( \phi ^{\dagger }\eta -\eta ^{\dagger }\phi
\right) , \notag\\
p_{1} &=&2\func{Re}\left( \varphi \rho \right) ,\ \hspace{0.5cm}\hspace{0.5cm%
}p_{2}=2\func{Im}\left( \varphi \rho \right) ,\ \hspace{0.5cm}\hspace{0.5cm}%
w_{1}=2\func{Re}\left( \rho \zeta \right) ,\ \hspace{0.5cm}\hspace{0.5cm}%
w_{2}=2\func{Im}\left( \rho \zeta \right) , \notag\\
o_{1} &=&2\func{Re}\left( \rho ^{2}\right) ,\ \hspace{0.5cm}\hspace{0.5cm}%
o_{2}=2\func{Im}\left( \rho ^{2}\right) ,\ \hspace{0.5cm}\hspace{0.5cm}%
o_{3}=2\func{Re}\left( \sigma ^{2}\right) ,\ \hspace{0.5cm}\hspace{0.5cm}%
o_{4}=2\func{Im}\left( \sigma ^{2}\right).
\end{eqnarray}
The quartic terms of the scalar potential can be rewritten as follows
\begin{eqnarray}
V_{4} &=&\left( \sqrt{\lambda _{1}}a-\sqrt{\lambda _{2}}b\right) ^{2}+\left( 
\sqrt{\lambda _{1}}a-\sqrt{\lambda _{4}}c\right) ^{2}+\left( \sqrt{\lambda
_{1}}a-\sqrt{\lambda _{5}}d\right) ^{2}+\left( \sqrt{\lambda _{1}}a-\sqrt{%
\lambda _{6}}e\right) ^{2} \notag\\
&&+\left( \sqrt{\lambda _{1}}a-\sqrt{\lambda _{7}}f\right) ^{2}+\left( \sqrt{%
\lambda _{2}}b-\sqrt{\lambda _{4}}c\right) ^{2}+\left( \sqrt{\lambda _{2}}b-%
\sqrt{\lambda _{5}}d\right) ^{2}+\left( \sqrt{\lambda _{2}}b-\sqrt{\lambda
_{6}}e\right) ^{2} \notag\\
&&+\left( \sqrt{\lambda _{2}}b-\sqrt{\lambda _{7}}f\right) ^{2}+\left( \sqrt{%
\lambda _{4}}c-\sqrt{\lambda _{5}}d\right) ^{2}+\left( \sqrt{\lambda _{4}}c-%
\sqrt{\lambda _{6}}e\right) ^{2}+\left( \sqrt{\lambda _{4}}c-\sqrt{\lambda
_{7}}f\right) ^{2} \notag\\
&&+\left( \sqrt{\lambda _{5}}d-\sqrt{\lambda _{6}}e\right) ^{2}+\left( \sqrt{%
\lambda _{5}}d-\sqrt{\lambda _{7}}f\right) ^{2}+\left( \sqrt{\lambda _{6}}e-%
\sqrt{\lambda _{7}}f\right) ^{2} \notag\\
&&-4\left( \lambda _{1}a^{2}+\lambda _{2}b^{2}+\lambda _{4}c^{2}+\lambda
_{5}d+\lambda _{6}e+\lambda _{7}f^{2}\right) +\lambda _{17}\left(
r^{2}+s^{2}\right)  \notag\\
&&+\left( \lambda _{3}+2\sqrt{\lambda _{1}\lambda _{2}}\right) ab+\left(
\lambda _{16}+2\sqrt{\lambda _{1}\lambda _{4}}\right) ac+\left( \lambda
_{18}+2\sqrt{\lambda _{1}\lambda _{5}}\right) ad \notag\\
&&+\left( \lambda _{19}+2\sqrt{\lambda _{1}\lambda _{6}}\right) ae+\left(
\lambda _{20}+2\sqrt{\lambda _{1}\lambda _{7}}\right) af+\left( \lambda
_{21}+2\sqrt{\lambda _{2}\lambda _{4}}\right) bc \notag\\
&&+\left( \lambda _{22}+2\sqrt{\lambda _{2}\lambda _{5}}\right) bd+\left(
\lambda _{23}+2\sqrt{\lambda _{2}\lambda _{6}}\right) be+\left( \lambda
_{24}+2\sqrt{\lambda _{2}\lambda _{7}}\right) bf \notag\\
&&+\left( \lambda _{16}+2\sqrt{\lambda _{4}\lambda _{5}}\right) cd+\left(
\lambda _{9}+2\sqrt{\lambda _{4}\lambda _{6}}\right) ce+\left( \lambda
_{10}+2\sqrt{\lambda _{4}\lambda _{7}}\right) cf \notag\\
&&+\left( \lambda _{11}+2\sqrt{\lambda _{5}\lambda _{6}}\right) de+\left(
\lambda _{12}+2\sqrt{\lambda _{5}\lambda _{7}}\right) df+\left( \lambda
_{13}+2\sqrt{\lambda _{6}\lambda _{7}}\right) ef \notag\\
&&+\frac{1}{8}\lambda _{14}\left( p_{1}o_{1}-p_{2}o_{2}\right) +\frac{1}{8}%
\lambda _{15}\left( w_{1}o_{3}-w_{2}o_{4}\right).
\end{eqnarray}
For the analysis of the tree-level stability of the scalar potential, we follow the method described in Refs.~\cite{Maniatis:2006fs, Bhattacharyya:2015nca}, we find the following tree-level stability conditions of the scalar potential
\begin{eqnarray}
\lambda _{1} &\geq &0,\hspace{0.5cm}\lambda _{2}\geq 0,\hspace{0.5cm}\lambda
_{4}\geq 0,\hspace{0.5cm}\lambda _{5}\geq 0,\hspace{0.5cm}\lambda _{6}\geq 0,%
\hspace{0.5cm} \nonumber\\
\lambda _{7} &\geq &0,\hspace{0.5cm}\lambda _{17}\geq 0,\hspace{0.5cm}%
\lambda _{14}\geq 0,\hspace{0.5cm}\lambda _{15}\geq 0, \nonumber\\
\lambda _{3}+2\sqrt{\lambda _{1}\lambda _{2}} &\geq &0,\hspace{0.5cm}\hspace{%
0.5cm}\lambda _{16}+2\sqrt{\lambda _{1}\lambda _{4}}\geq 0,\hspace{0.5cm}%
\hspace{0.5cm}\lambda _{18}+2\sqrt{\lambda _{1}\lambda _{5}}\geq 0, \nonumber\\
\lambda _{19}+2\sqrt{\lambda _{1}\lambda _{6}} &\geq &0,\hspace{0.5cm}%
\hspace{0.5cm}\lambda _{20}+2\sqrt{\lambda _{1}\lambda _{7}}\geq 0,\hspace{%
0.5cm}\hspace{0.5cm}\lambda _{21}+2\sqrt{\lambda _{2}\lambda _{4}}\geq 0, \nonumber\\
\lambda _{22}+2\sqrt{\lambda _{2}\lambda _{5}} &\geq &0,\hspace{0.5cm}%
\hspace{0.5cm}\lambda _{23}+2\sqrt{\lambda _{2}\lambda _{6}}\geq 0,\hspace{%
0.5cm}\hspace{0.5cm}\lambda _{24}+2\sqrt{\lambda _{2}\lambda _{7}}\geq 0, \nonumber\\
\lambda _{16}+2\sqrt{\lambda _{4}\lambda _{5}} &\geq &0,\hspace{0.5cm}%
\hspace{0.5cm}\lambda _{9}+2\sqrt{\lambda _{4}\lambda _{6}}\geq 0,\hspace{%
0.5cm}\hspace{0.5cm}\lambda _{10}+2\sqrt{\lambda _{4}\lambda _{7}}\geq 0 \nonumber\\
\lambda _{8}+2\sqrt{\lambda _{4}\lambda _{5}} &\geq &0,\hspace{0.5cm}\hspace{%
0.5cm}\lambda _{12}+2\sqrt{\lambda _{5}\lambda _{7}}\geq 0,\hspace{0.5cm}%
\hspace{0.5cm}\lambda _{13}+2\sqrt{\lambda _{6}\lambda _{7}}\geq 0.
\end{eqnarray}

\bibliographystyle{utphys}
\bibliography{biblio}

\end{document}